\documentclass[aps,superscriptaddress,amsmath,amssymb,floatfix,twocolumn,showpacs,amsfonts,longbibliography]{revtex4-1}
\usepackage{times}
\usepackage[varg]{txfonts}
\usepackage{textcomp}
\usepackage{graphicx}
\usepackage{subfigure}
\usepackage{tabu}
\usepackage{color}
\usepackage[colorlinks=true,citecolor=blue,urlcolor=blue,linkcolor=blue,hyperindex]{hyperref}
\usepackage{braket}
\usepackage{overpic}
\usepackage{amssymb}
\allowdisplaybreaks

\begin{document}
	\title{XXZ-Ising model on the triangular Kagome lattice with spin-1 on the decorated trimers}
	\author{Chengkang Zhou}
	\affiliation{State Key Laboratory of Optoelectronic Material and Technologies, School of Physics, Sun Yat-sen University, Guangzhou 510275, China}

	\author{Yuanwei Feng}
	\affiliation{State Key Laboratory of Optoelectronic Material and Technologies, School of Physics, Sun Yat-sen University, Guangzhou 510275, China}
		
	\author{Jiawei Ruan}
	\affiliation{State Key Laboratory of Optoelectronic Material and Technologies, School of Physics, Sun Yat-sen University, Guangzhou 510275, China}
	\affiliation{National Laboratory of Solid State Microstructures, School of Physics and Collaborative Innovation Center of Advanced Microstructures, Nanjing University, Nanjing 210093, China.}	
		
	\author{Dao-Xin Yao}
	\email[Corresponding author:]{yaodaox@mail.sysu.edu.cn}
	\affiliation{State Key Laboratory of Optoelectronic Material and Technologies, School of Physics, Sun Yat-sen University, Guangzhou 510275, China}
	
	\date{\today}
	
	\begin{abstract}
	We consider the triangular Kagome XXZ-Ising model (TKL XXZ-Ising model) formed by inserting small triangles ("a-trimers") with XXZ spin-1 inside the triangles of the Kagome lattice ("b-trimers"). It is a novel mixed spin system and can be solved exactly by transforming into the Kagome lattice with the general transformation method for decorated spin systems. In the absence of an external field, we integrate out the quantum spins of the a-trimers and map the TKL model to the Kagome Ising model exactly. We obtain the full phase diagram and their zero-temperature entropies (e.g. $s_{max}=5.48895$ per unit cell is given for the phase with the maximum entropy). When an external field is applied, 20 phases are found due to the quantum fluctuations of a-trimers. Moreover, the high spins in the a-trimers can lead to a stable quantized growth of the magnetization process in the Heisenberg limit.
	
	\end{abstract}
    \maketitle
	
	\section{INTRODUCTION}
	\label{INTRODUCTION}	
	Introducing quantum fluctuations into a classical model has both fundamental and practical importance for finding new quantum phases at low temperature. Especially, it can lead to a multitude of new quantum phases and nontrivial phase transitions in frustrated systems with a large ground state degeneracy, for example, the tetramer-dimer and dimer-monomer phases in the frustrated Heisenberg diamond chain\cite{takano1996ground}, Kagome loop gas in the triangle Kagome lattice\cite{yao2008xxz}. It is known that these unusual phases are the result of the interplay between quantum fluctuations and geometric frustrations\cite{lhuillier2001frustrated}. Since the degenerate states in such systems have the same energy levels and all perturbations are singular, any linear combination of the classically degenerate states is a candidate for new quantum ground states\cite{moessner1999two-dimensional}. Moreover, this effect plays an important role in frustrated mixed spin systems, which includes both spin-$1/2$ and higher spins. Sorts of classical degenerate states exist in such spin system.
	
\begin{figure}
		\includegraphics[height=6cm]{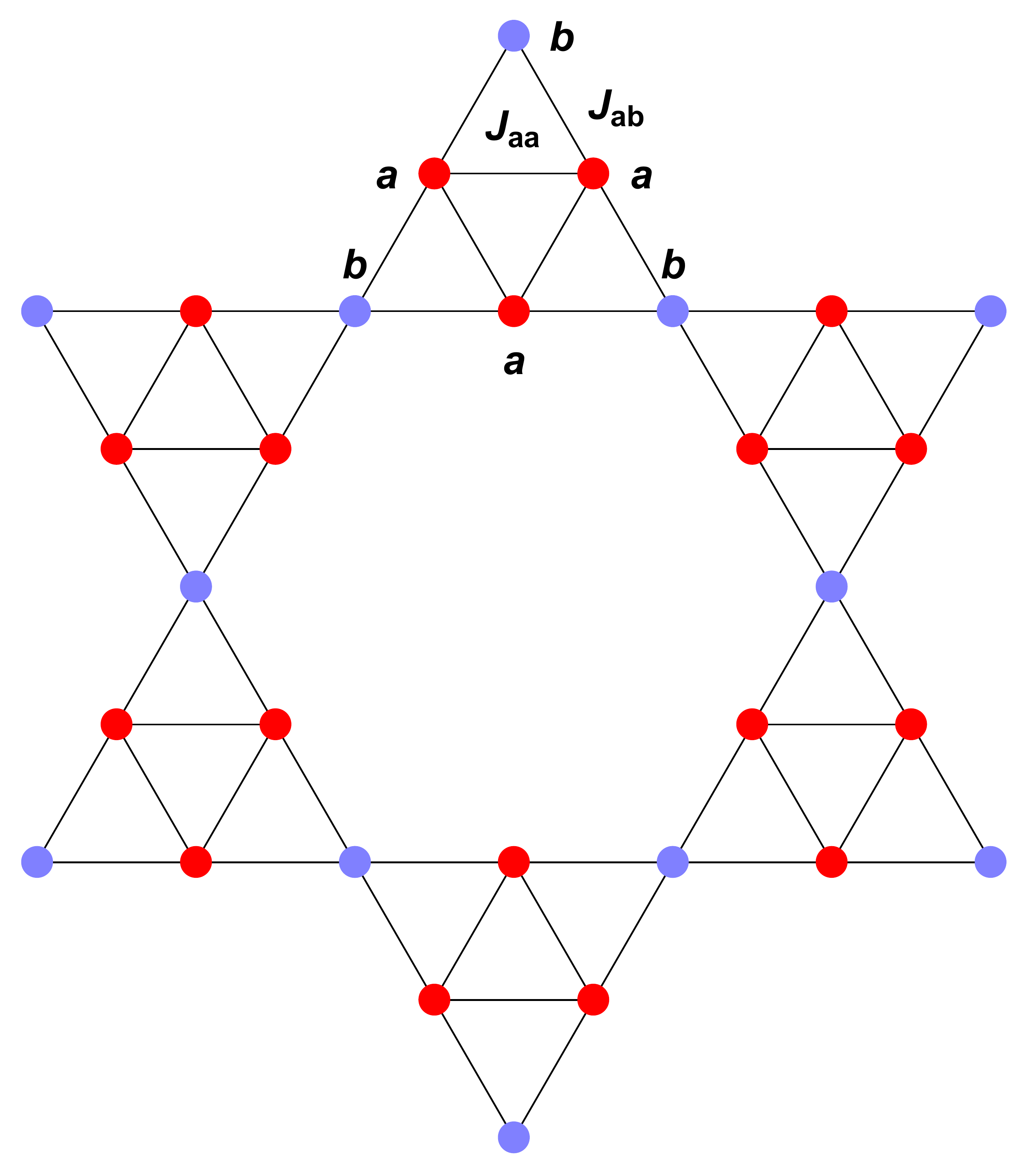}
		\caption{The XXZ-Ising model of triangular Kagome lattice (TKL) formed by introducing small triangles ("a-trimers", red (dark)) with XXZ spins on the Kagome lattice ("b-trimers", blue (light)) with Ising spins.}
		\label{Fig1}
	\end{figure}

	Theoretical interest in mixed spin systems is increasing in recent years. Most of the mixed spin systems are constructed by inserting high spin decorated parts in the standard Ising spin systems. For instance, the high spin decorated parts in a diamond chain exhibits an outstanding magnetization properties\cite{canova2009exact}. Among them, the triangle Kagome lattice (TKL) is a typical structure formed by inserting small triangles into the large triangles in the Kagome lattice (see Fig.\ref{Fig1}). It was found in $\mathrm{Cu}_9\mathrm{X}_2(\mathrm{cpa})_6\cdot \mathrm{xH}_2\mathrm{O}$ in the 1990s \cite{doi:10.1063/1.357006,doi:10.1080/10587259308054973,ateca1995the,doi:10.1143/JPSJ.81.053703}. Previous researches have revealed that the TKL XXZ-Ising model with the spin-$1/2$ on the decorated trimers can be solved exactly\cite{moessner2000two-dimensional,moessner2001ising,moessner2003theory,loh2008dimers,zheng2005exact,yamamoto2014quantum,doi:10.1063/1.357006,doi:10.1080/10587259308054973,ateca1995the,doi:10.1143/JPSJ.81.053703,yao2008xxz,loh2007thermodynamics,strecka2008exact}. However, in the presence of mixed spin case, the TKL XXZ-Ising model calls for farther investigations.
	
	One of the most important way to study decorated spin system is the general transformation method. It was first introduced by Fisher in the 1950s\cite{fisher1959transformations} and developed in recent years\cite{rojas2009generalized,rojas2009generalized,strecka2010generalized}. And it has been widely applied in studying the decorated spin system, in both one\cite{antonosyan2008exactly,galisova2013magnetic,ananikian2014magnetization} and two\cite{0305-4470-35-16-103,hanks2013exact,chunfeng2006a,strecka2013order-from-disorder} dimensions. With this method, the TKL XXZ-Ising model remains solvable when changing the decorated parts with higher spins, which make it serve as an ideal candidate for observing the effects of quantum fluctuations in the mixed spin systems with geometric frustrations.
	
	In this paper, we investigate the TKL XXZ-Ising model decorated by the spin-$1$ trimers (spin-$1$ TKL model). By comparing the pure spin-$1/2$ and the mixed spin cases, we give a picture of how the phase diagram of the TKL XXZ-Ising model evolves when the decorated spins turn higher.
	
	The rest of this paper is organized as follows. In Sec.\ref{MM}, we give the Hamiltonian of the TKL XXZ-Ising model decorated by the spin-$1$ trimers and introduce the transformation method. In Sec.\ref{Sec:ZF}, we discuss the zero temperature phase diagram without the external field and compare with the spin-$1/2$ TKL case. In Sec.\ref{Sec:FFZ}, we present the zero temperature phase diagram in the presence of a finite magnetic field and discuss the effect of the higher spin decorated parts. In Sec.\ref{Sec:CONCLUSION}, we present our final discussion and conclusion.

	\section{MODEL AND METHOD}
	\label{MM}
	
	The TKL XXZ-Ising model has two different kinds of sub-lattices, which are the a-trimers (the red triangle in Fig.\ref{Fig1})and the b-trimers (the blue triangle in Fig.\ref{Fig1}). In this model, we consider the exchange couplings between the a-spins (the spins in the a-trimers) are of the XXZ type and the couplings between the neighboring a-spins and b-spins (the spins in the b-trimers) are of the Ising type. The Hamiltonian of the spin-1 TKL model is given by
	\begin{equation}
	\begin{aligned}
	H=&-J_{a}^{xy}\sum_{\mathrm{ai},\mathrm{aj}\in a}(S_{\mathrm{ai}}^xS_{\mathrm{aj}}^x+S_{\mathrm{ai}}^yS_{\mathrm{aj}}^y)-J_{a}^z\sum_{\mathrm{ai},\mathrm{aj}\in a} S_{\mathrm{ai}}^zS_{\mathrm{aj}}^z
	\\&-J_{ab}^{z}\sum_{\mathrm{ai}\in a,\mathrm{bi}\in b}S_{\mathrm{ai}}^zS_{\mathrm{bi}}^z-h\sum_{\mathrm{ai}\in a} S_{\mathrm{ai}}^z-h\sum_{\mathrm{bi}\in b} S_{\mathrm{bi}}^z.
	\label{Ham01}
	\end{aligned}
	\end{equation}
	
\begin{figure}
		\includegraphics[height=3cm]{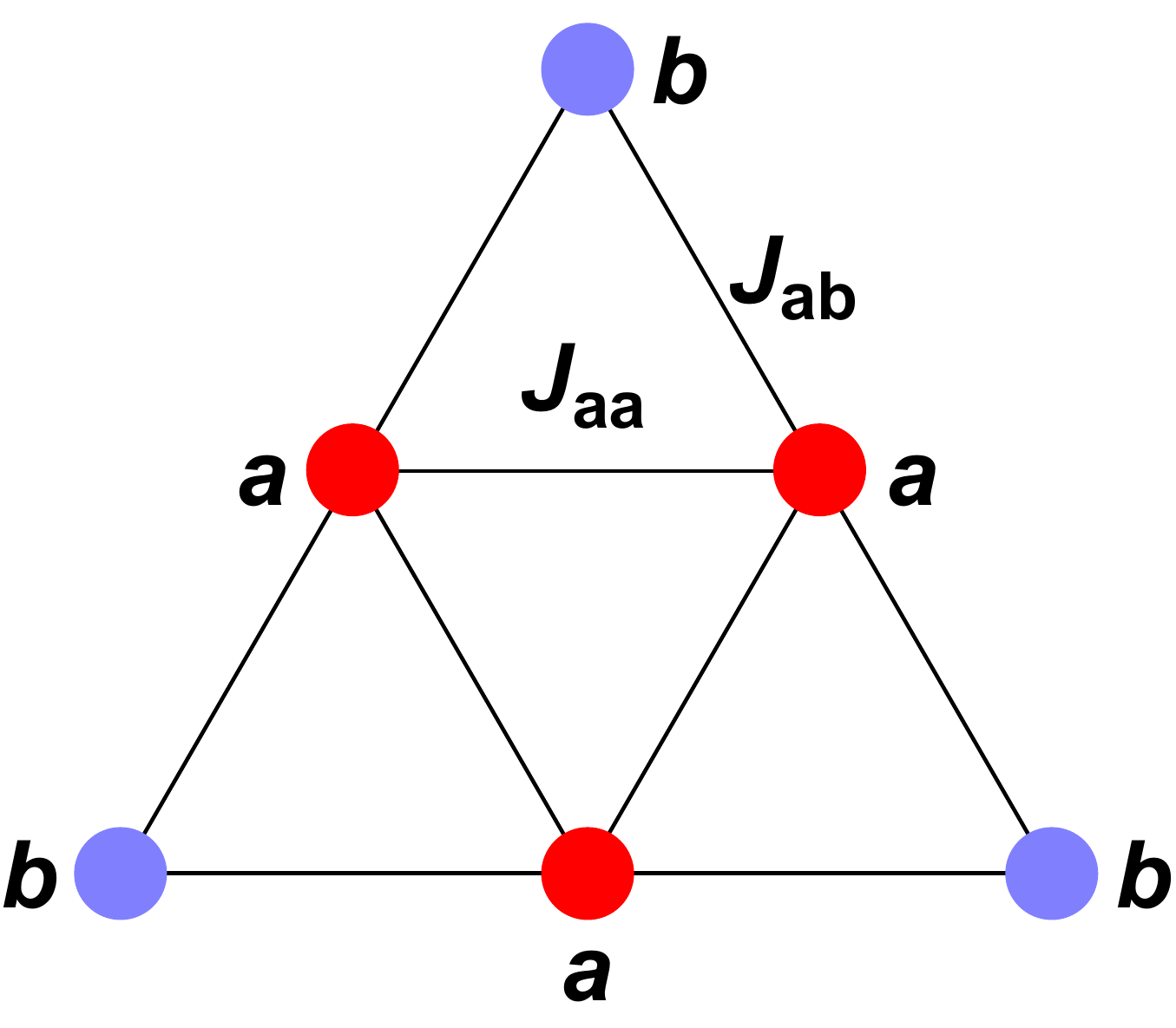}
		\caption{A hexamer contains three spins on the a-sublattice and three spins on the b-sublattice. Each b-spin is shared by two hexamers}
		\label{Fig2}
	\end{figure}

	In which, the spins carry $S=1/2$ and $1$ for the b-spins and the a-spins respectively. $J_{ab}^{z}$ denotes the Ising coupling between the a-spins and the b-spins. $J_{a}^z$ ($J_{a}^{xy}$) is the coupling of the a-spins in the $z$ ($x$ and $y$) direction respectively. $h$ is the applied external field. We set $|J_{ab}^{z}|$ to be unit of energy in the following analysis. The Hamiltonian Eq.\ref{Ham01} can be written as a sum of the hexamers (see Fig.\ref{Fig2}), which is
	\begin{equation}
	\begin{aligned}
	H=\sum_{n}H_n,
	\end{aligned}
	\end{equation}
	\begin{equation}
	\begin{aligned}
	H_n=&-J_{a}^{xy}\sum_{\mathrm{ai},\mathrm{aj}\in a}(S_{\mathrm{ai}}^xS_{\mathrm{aj}}^x+S_{\mathrm{ai}}^yS_{\mathrm{aj}}^y)-J_{a}^z\sum_{\mathrm{ai},\mathrm{aj}\in a} S_{\mathrm{ai}}^zS_{\mathrm{aj}}^z
	\\&-J_{ab}^{z}\sum_{\mathrm{ai}\in a,\mathrm{bi}\in b}S_{\mathrm{ai}}^zS_{\mathrm{bi}}^z-h\sum_{\mathrm{ai}\in a} S_{\mathrm{ai}}^z-\frac{h}{2}\sum_{\mathrm{bi}\in b} S_{\mathrm{bi}}^z.
	\label{Ham_each_hexamer}
	\end{aligned}
	\end{equation}
		
	Note that the Hamiltonian of each hexamer (Eq.\ref{Ham_each_hexamer}) commutes with each other. And each a-spin appears just in one hexamer. Therefore, the eigenvector of each hexamer has the form as 
		
	\begin{equation}
	\begin{aligned}
	\ket{\mathrm{hexamer}}=\ket{S_{\mathrm{bi}}^{z}, S_{\mathrm{bj}}^{z}, S_{\mathrm{bk}}^{z}} \otimes \ket{S_{\mathrm{ai}}^{z}, S_{\mathrm{aj}}^{z}, S_{\mathrm{ak}}^{z}}.
	\end{aligned}
	\label{eigenvector}
	\end{equation} 
	
	Hence, the decorated trimers are localized and it is reasonable to trace over all the a-spins for each hexamer. The partition function of each hexamer is given by
	
	 \begin{equation}
	 \begin{aligned}
	 Z(S_{\mathrm{b1}}^z,S_{\mathrm{b2}}^z,S_{\mathrm{b3}}^z)= \mathrm{Tr} e^{-\beta H_{n}(\hat{S_{\mathrm{ai}}},S_{\mathrm{bi}}^z)}.\\
	 \end{aligned}
	 \end{equation}

	The trace can be evaluated by diagonalizing Hamiltonian Eq.\ref{Ham_each_hexamer} for each configuration of the enclosing b-spins. When considering the $C_3$ symmetry in the spin-$1$ TKL model ($Z(\downarrow\uparrow\uparrow,h)=Z(\uparrow\downarrow\uparrow,h)=Z(\uparrow\uparrow\downarrow,h)$), there are only four different configurations, which are $Z(\uparrow\uparrow\uparrow,h)$, $Z(\downarrow\downarrow\downarrow,h)$, $Z(\downarrow\uparrow\uparrow,h)$ and $Z(\uparrow\downarrow\downarrow,h)$. We give their explicit form in Appendix.\ref{Explicit Form}. 
	
\begin{figure}
		\includegraphics[height=3cm]{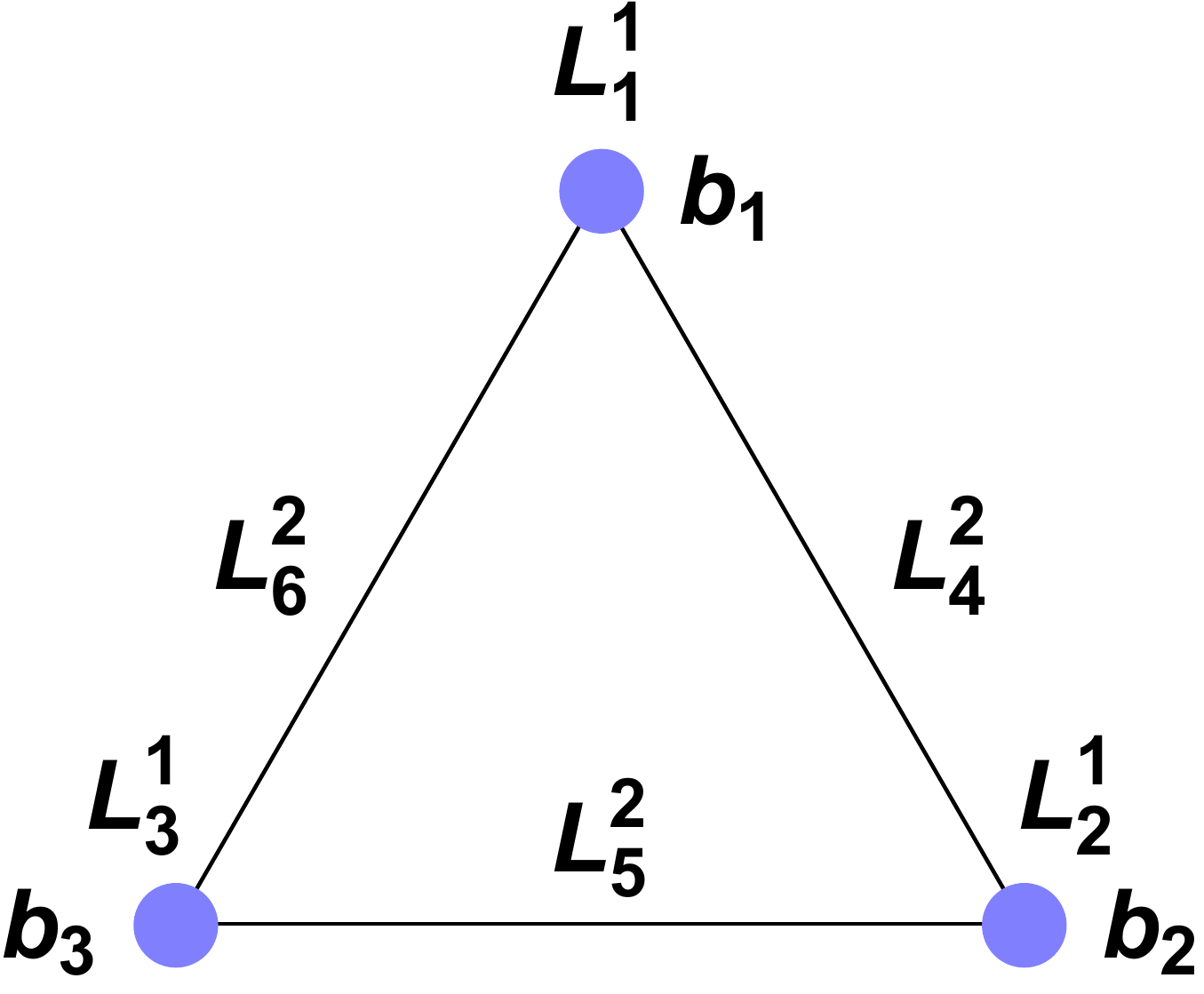}
		\caption{The effective trimer after transformation. $L_0^0$ corresponds to the coefficient of effective background energy, $L_1^1$, $L_2^1$ and $L_3^1$ correspond to the coefficients of one spin coupling, $L_4^2$, $L_5^2$ and $L_6^2$ correspond to the coefficients of two spins coupling, $L_7^3$ correspond to the coefficients of three spin coupling (not shown in Fig.\ref{Fig3}).}
		\label{Fig3}
	\end{figure}

	Since it is just a function of the b-spins, it is available to transform the hexamer into an effective trimer, in which only contains the b-spins (see Fig.\ref{Fig3}). According to the general transformation method for the decorated spin systems\cite{rojas2009generalized,strecka2010generalized}, the Hamiltonian of the effective trimer can be assumed as	
	\begin{equation}
	\begin{aligned}
	H_{n}'=&-L_0^0-L_1^1\sigma_{\mathrm{b1}}^z-L_2^1\sigma_{\mathrm{b2}}^z-L_3^1\sigma_{\mathrm{b3}}^z\\
	&-L_4^2\sigma_{\mathrm{b1}}^z\sigma_{\mathrm{b2}}^z-L_5^2\sigma_{\mathrm{b2}}^z\sigma_{\mathrm{b3}}^z-L_6^2\sigma_{\mathrm{b1}}^z\sigma_{\mathrm{b3}}^z-L_7^3\sigma_{\mathrm{b1}}^z\sigma_{\mathrm{b2}}^z\sigma_{\mathrm{b3}}^z.
	\end{aligned}
	\end{equation}
	
	Here we use $\sigma_{\mathrm{bi}}^z=\pm1$ for consistency with the Ising model literature. $L_0^0$ stands for the parameter of the effective background energy. $L_1^1$, $L_2^1$ and $L_3^1$ represent the coefficients of each b-spin coupling with the effective fields. $L_4^2$, $L_5^2$ and $L_6^2$ are the coefficients of two b-spins effective couplings. $L_7^3$ carries the coefficient of three b-spins effective coupling. Since the effective trimer is classical, the partition function of each b-spins configuration can be written as
	\begin{equation}
	\begin{aligned}
	Z(\sigma_{\mathrm{b1}}^z,\sigma_{\mathrm{b2}}^z,\sigma_{\mathrm{b3}}^z)=\exp[{-\beta H_{n}'(\sigma_{\mathrm{b1}}^z,\sigma_{\mathrm{b2}}^z,\sigma_{\mathrm{b3}}^z)}].\\
	\end{aligned}
	\end{equation}
	
	To keep the partition function,  $Z(S_{\mathrm{b1}}^z,S_{\mathrm{b2}}^z,S_{\mathrm{b3}}^z)=Z(\sigma_{\mathrm{b1}}^z,\sigma_{\mathrm{b2}}^z,\sigma_{\mathrm{b3}}^z)$ when they share the same b-spin configuration. Hence,  
	
	\begin{equation}
	\begin{aligned}
	Z(S_{\mathrm{b1}}^z,S_{\mathrm{b2}}^z,S_{\mathrm{b3}}^z)=\exp[{-\beta H_{n}'(\sigma_{\mathrm{b1}}^z,\sigma_{\mathrm{b2}}^z,\sigma_{\mathrm{b3}}^z)}].\\
	\end{aligned}
	\label{Z}
	\end{equation}
	
	As a result, the effective couplings can be expressed by the partition functions of each hexamer. To give their formulas, it is more convenient to take the logarithm of both sides of Eq.\ref{Z}\cite{rojas2009generalized},
	
	\begin{equation}
	\ln[Z]=V^{\frac{1}{2}}\otimes V^{\frac{1}{2}}\otimes V^{\frac{1}{2}} \beta L,\\
	\end{equation}
	
	where
	\begin{equation}
	\ln[Z]=\left(
	\begin{array}{c}
	\ln[Z(\uparrow\uparrow\uparrow)]\\
	\ln[Z(\downarrow\uparrow\uparrow)]\\
	\ln[Z(\uparrow\downarrow\uparrow)]\\
	\ln[Z(\downarrow\downarrow\uparrow)]\\
	\ln[Z(\uparrow\uparrow\downarrow)]\\
	\ln[Z(\downarrow\uparrow\downarrow)]\\
	\ln[Z(\uparrow\downarrow\downarrow)]\\
	\ln[Z(\downarrow\downarrow\downarrow)]\\
	\end{array}
	\right) , \beta L=\left(
	\begin{array}{c}
	\beta L_0^0\\
	\beta L_3^1\\
	\beta L_2^1\\
	\beta L_5^2\\
	\beta L_1^1\\
	\beta L_6^2\\
	\beta L_4^2\\
	\beta L_7^3\\
	\end{array}
	\right),
	\end{equation}
	
	\begin{equation}
	V^{\frac{1}{2}}=\left(
	\begin{array}{cc}
	1 & 1 \\
	1 & -1 \\
	\end{array}
	\right).
	\end{equation}
	
	With $(\uparrow\uparrow\uparrow)$, for instance, represents one possible configuration of the b-spins. Finally, the effective couplings can be expressed as
	
	\begin{equation}
	\beta L=\left(
	\begin{array}{cccccccc}
	1 & 1 & 1 & 1 & 1 & 1 & 1 & 1 \\
	1 & -1 & 1 & -1 & 1 & -1 & 1 & -1 \\
	1 & 1 & -1 & -1 & 1 & 1 & -1 & -1 \\
	1 & -1 & -1 & 1 & 1 & -1 & -1 & 1 \\
	1 & 1 & 1 & 1 & -1 & -1 & -1 & -1 \\
	1 & -1 & 1 & -1 & -1 & 1 & -1 & 1 \\
	1 & 1 & -1 & -1 & -1 & -1 & 1 & 1 \\
	1 & -1 & -1 & 1 & -1 & 1 & 1 & -1 \\
	\end{array}
	\right)\frac{\ln[Z]}{8}.\\
	\end{equation}
	
	In the TKL model, we can simplify the effective couplings with its $C_3$ symmetry, which is
	\begin{equation}
	\begin{aligned}
	\ln(Z_{a})&=\beta L_0^0\\&=\ln\left[[Z(\uparrow\uparrow\uparrow,h)Z(\downarrow\downarrow\downarrow,h)]^{\frac{1}{8}}[Z(\downarrow\uparrow\uparrow,h)Z(\uparrow\downarrow\downarrow,h)]^{\frac{3}{8}}\right],
	\label{Za}
	\end{aligned}
	\end{equation}
	\begin{equation}
	\begin{aligned}
	\beta h_{b}&=\beta L_1^1=\beta L_2^1=\beta L_3^1\\&=\frac{1}{8}\ln\left[\frac{Z(\uparrow\uparrow\uparrow,h)Z(\downarrow\uparrow\uparrow,h)}{Z(\downarrow\downarrow\downarrow,h)Z(\uparrow\downarrow\downarrow,h)}\right],
	\label{hb}
	\end{aligned}
	\end{equation}
	\begin{equation}
	\begin{aligned}
	\beta J_{bb}&=\beta L_4^2=\beta L_5^2=\beta L_6^2\\&=\frac{1}{8}\ln\left[\frac{Z(\uparrow\uparrow\uparrow,h)Z(\downarrow\downarrow\downarrow,h)}{Z(\downarrow\uparrow\uparrow,h)Z(\uparrow\downarrow\downarrow,h)}\right],
	\label{bjbb}
	\end{aligned}
	\end{equation}
	\begin{equation}
	\begin{aligned}
	\beta J_{bbb}=\beta L_7^3=\frac{1}{8}\ln\left[ \frac{Z(\uparrow\uparrow\uparrow,h)Z(\uparrow\downarrow\downarrow,h)^3}{Z(\downarrow\downarrow\downarrow,h)Z(\downarrow\uparrow\uparrow,h)^3}\right].
	\label{bjbbb}
	\end{aligned}
	\end{equation}
	
	It is worth noting that $h_{b}$ should be doubled when considering the whole model since each b-spin is shared by two hexamers. Eventually, the Hamiltonian of the effective trimer becomes
	\begin{equation}
	\begin{aligned}
	H_{n}'(\sigma_{\mathrm{bi}}^{z})=&-\ln(Z_{a})/\beta-h_{b}\sum_{\mathrm{bi}\in b}\sigma_{\mathrm{bi}}^z-J_{bb}\sum_{<\mathrm{bi},\mathrm{bj}>}\sigma_{\mathrm{bi}}^z\sigma_{\mathrm{bj}}^z\\&-J_{bbb}\sum_{<\mathrm{bi},\mathrm{bj},\mathrm{bk}>}\sigma_{\mathrm{bi}}^z\sigma_{\mathrm{bj}}^z\sigma_{\mathrm{bk}}^z.
	\end{aligned}
	\label{eqe}
	\end{equation}
	
	With Eq.\ref{eqe}, the spin-$1$ TKL model can be exactly mapped to the classical Kagome model with an extra three-spin coupling. Consequently, we can obtain the zero-temperature ground state of the b-spins in the usual manner, which is searching for the lowest energy state of each unit cell. We present the phase diagram and discuss the effect of the interplay between the quantum fluctuations and the geometric frustrations in the spin-$1$ TKL model in the following sections.
	
	\section{ZERO FIELD}
	\label{Sec:ZF}
	
	\subsection{Mapping to the Kagome Ising model}
	Due to the time-reversal symmetry ($Z(\uparrow\uparrow\uparrow,h)=Z(\downarrow\downarrow\downarrow,-h)$, $Z(\downarrow\uparrow\uparrow,h)=Z(\uparrow\downarrow\downarrow,-h)$), the effective couplings of the spin-$1$ TKL model can be farther simplified as
	\begin{equation}
	\begin{aligned}
	H_{n}'=&-\ln(Z_{a})/\beta-J_{bb}\sum_{\mathrm{bi},\mathrm{bj}\in b}\sigma_{\mathrm{bi}}^z\sigma_{\mathrm{bj}}^z,
	\end{aligned}
	\end{equation}
	in which
	\begin{equation}
	\begin{aligned}
	Z_{a}=Z(\uparrow\uparrow\uparrow)^{\frac{1}{4}}Z(\downarrow\uparrow\uparrow)^{\frac{3}{4}},
	\end{aligned}
	\end{equation}
	\begin{equation}
	\begin{aligned}
	\beta J_{bb}=\frac{1}{4}\ln\left[\frac{Z(\uparrow\uparrow\uparrow)}{Z(\downarrow\uparrow\uparrow)}\right],
	\end{aligned}
	\end{equation}
	\begin{equation}
	\begin{aligned}
	\beta h_{b}&=\beta J_{bbb}=0.
	\end{aligned}
	\end{equation}
	
	After the transformation, the TKL model is mapped to the Kagome Ising model exactly, in which all the parameters ($J_{bb}$ and $Z_{a}$) are the functions of the original couplings ($J_{ab}^{z}$, $J_{a}^z$ and $J_{a}^{xy}$).

	\subsection{Free energy and entropy}
	
\begin{figure}[t]
		\centering
			{
			\includegraphics[height=5.5cm]{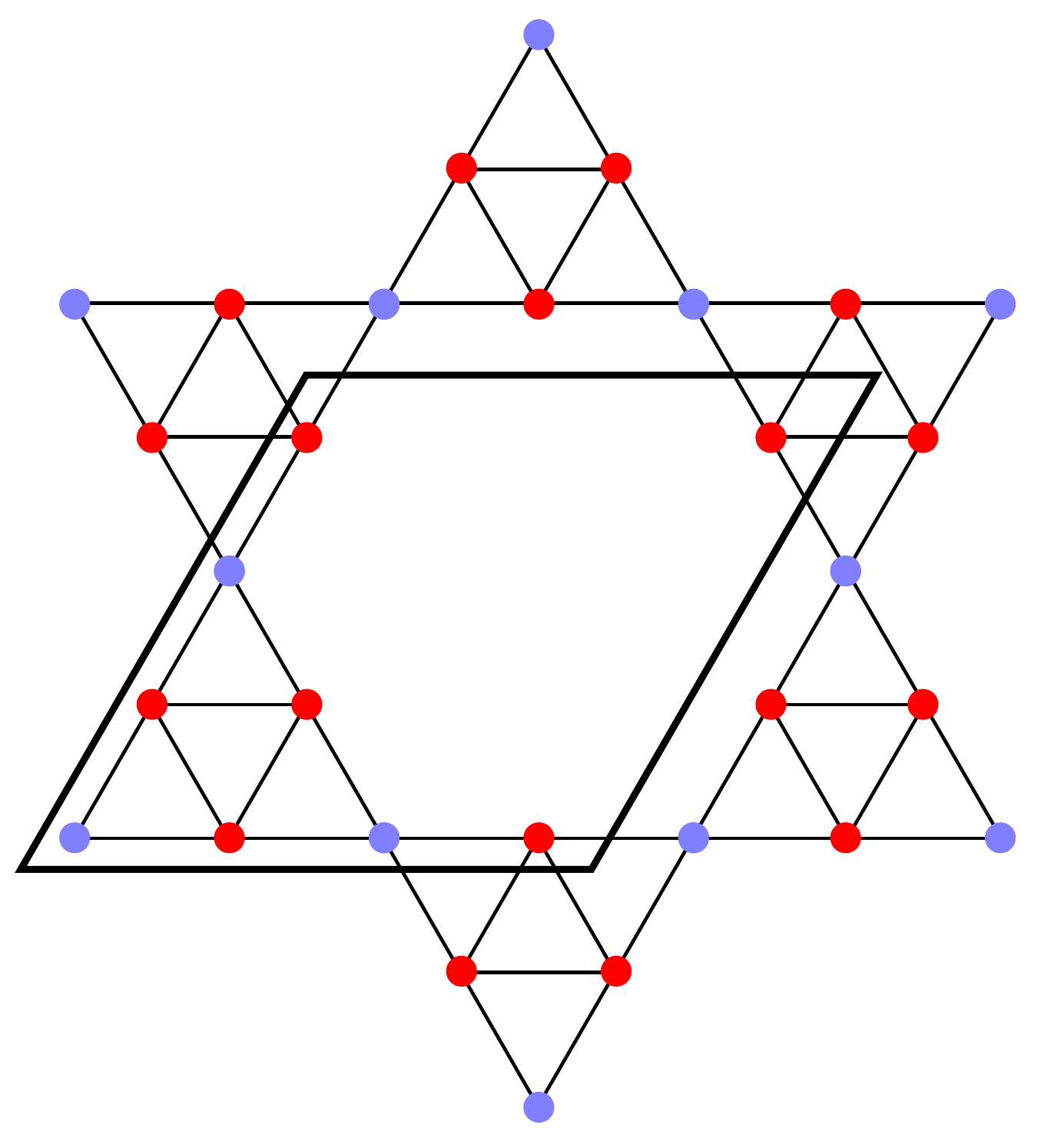}}
		\caption{A unit cell is shown as the blocked area.}
		\label{FigUC}
	\end{figure}

	Since the spin-$1$ TKL model has been mapped to the kagome Ising model, it is rational to compute the partition function, free energy, internal energy and entropy per unit cell of the spin-$1$ TKL model by applying the exact solution of the kagome Ising model\cite{kano1953antiferromagnetism.,syozi1951statistics}. For convenience, we define the $f$ as $f=\ln(Z)$ associating with free energy. Then in the spin-$1$ TKL model, it can be written as a sum of $f$ from the effective Kagome Ising model and from the a-trimers,
	
	\begin{equation}
	\begin{aligned}
	f_{\mathrm{TKL}}(J_{a}^{xy},J_{a}^{z},J_{ab}^{z})=f_{b}(\beta J_{bb})+2f_{a},
	\end{aligned}
	\label{f}
	\end{equation}
	in which $f_{a}=\ln(Z_{a})$. The factor 2 in Eq.\ref{f} comes from the fact that one unit cell of the spin-$1$ TKL model contains one b-trimer and two a-trimers (see Fig.\ref{FigUC}), which is different from the hexamers.
	
	The internal energy per unit cell is
	\begin{equation}
	\begin{aligned}
	u_{\mathrm{TKL}}&=-\frac{d{f_{\mathrm{TKL}}}}{d\beta}\\&=-\frac{d{f_{b}}}{d\beta}-2\frac{d{f_{a}}}{d\beta}\\&=u_{b}+u_{a},
	\end{aligned}
	\end{equation}
	where
	\begin{equation}
	\begin{aligned}
	u_{a}=-2\frac{d{f_{a}}}{d\beta}=\frac{u_{\uparrow\uparrow\uparrow}+3u_{\downarrow\uparrow\uparrow}}{2},
	\end{aligned}
	\end{equation}
	\begin{equation}
	\begin{aligned}
	u_{b}=-\frac{d{f_{b}}}{d\beta}=\frac{u_{\downarrow\uparrow\uparrow}-u_{\uparrow\uparrow\uparrow}}{4}u_{\mathrm{kag}}(\beta J_{bb}).
	\end{aligned}
	\end{equation}
	
	Here, we define $u_{\uparrow\uparrow\uparrow}=[E_{\uparrow\uparrow\uparrow}\exp(-\beta E_{\uparrow\uparrow\uparrow})]/Z$ for instance, which agrees with Ref.\cite{yao2008xxz}. And $u_{\mathrm{kag}}(\beta J_{bb})$ is the internal energy per unit cell of the effective Kagome Ising model. Finally, it becomes
	
	\begin{equation}
	\begin{aligned}
	u_{\mathrm{TKL}}=\frac{u_{\uparrow\uparrow\uparrow}+3u_{\downarrow\uparrow\uparrow}}{2}+\frac{u_{\downarrow\uparrow\uparrow}-u_{\uparrow\uparrow\uparrow}}{4}u_{\mathrm{kag}}(\beta J_{bb}).
	\end{aligned}
	\end{equation}
	
	The entropy per unit cell is
	\begin{equation}
	\begin{aligned}
	s_{\mathrm{TKL}}=f_{\mathrm{TKL}}+\beta u_{\mathrm{TKL}}.
	\end{aligned}
	\end{equation}
	
	Since $u_{\uparrow\uparrow\uparrow}$ and $u_{\downarrow\uparrow\uparrow}$ are dominated by the lowest energy of each hexamer at zero temperature, we can compute the free energy and entropy of the spin-$1$ TKL model with the exact solutions of the Kagome Ising model\cite{kano1953antiferromagnetism.,syozi1951statistics,Liebmann1986}. In the meanwhile, $\beta J_{bb}$ serves as the most important effective couplings since it is the decisive parameter in the Kagome Ising model.
	
	\begin{figure}[t]
		\centering
		{
				\includegraphics[height=5cm]{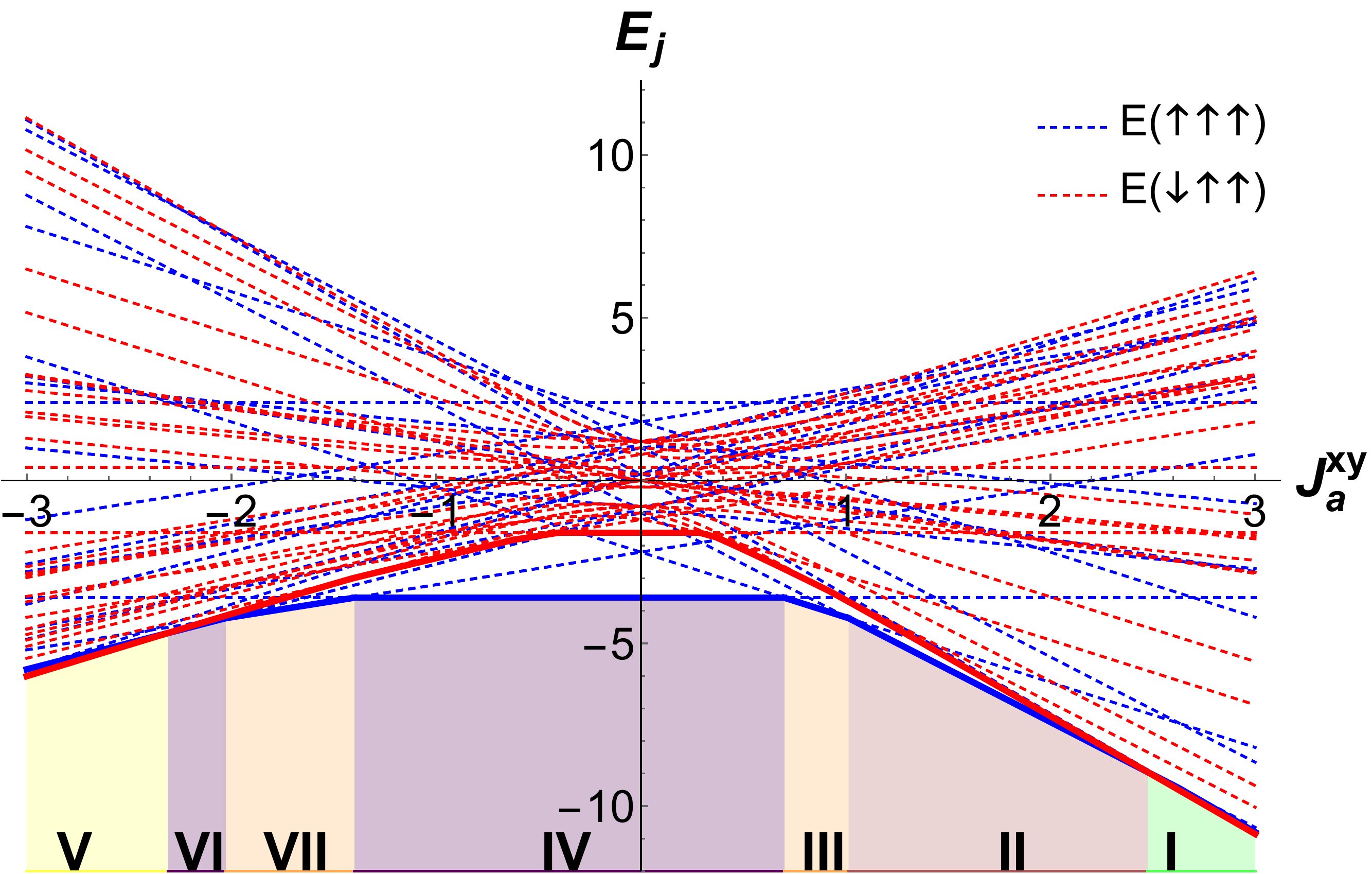}
		}
		\caption{The energy levels of each hexamer when the b-spin configuration is ($(\uparrow\uparrow\uparrow)$, dotted blue line) or ($(\downarrow\uparrow\uparrow)$, dotted red line) with $J^{z}_{a}=0.2$, $J^{z}_{ab}=-1$, $h=0$ and different $J^{xy}_{a}$ in Fig.\ref{FigE}. We also highlight the lowest ground state energy level of each b-spins configuration with the full lines and point out each phase with the same marks as in Fig.\ref{Fig4}.}
		\label{FigE}
	\end{figure}
	
	\subsection{Phase diagram at zero temperature and ground state properties}
	
	Since the Hilbert space of the hexamers can be divided into a-spins' space and b-spins' space (see Eq.\ref{eigenvector}), the best way to present its phase diagram is investigating the states of the b-spins and the a-spins respectively.
	
	For the b-spins, $\beta J_{bb}$ determines their behaviors. When $T\rightarrow0$, $\beta J_{bb}$ becomes
	\begin{equation}
	\begin{aligned}
	\beta J_{bb}=\frac{1}{4}\left[\ln\left(\frac{D_1}{D_{2}}\right)+\beta\left[E_0(\downarrow\uparrow\uparrow)-E_0(\uparrow\uparrow\uparrow)\right]\right].
	\end{aligned}
	\end{equation}	
	
	Here, $E_0(\uparrow\uparrow\uparrow)$($E_0(\downarrow\uparrow\uparrow)$) and $D_1$($D_2$) denote the ground state energy and the ground state degeneracy of each hexamer when the b-spins configuration is $(\uparrow\uparrow\uparrow)$($(\downarrow\uparrow\uparrow)$) at zero temperature respectively. We also define $D$ as the degeneracy of each hexamer in following discussion. Since $\beta\rightarrow\infty$ when $T\rightarrow0$, the sign of $\beta J_{bb}$ is determined by the competition between $E_0(\uparrow\uparrow\uparrow)$ and $E_0(\downarrow\uparrow\uparrow)$.
	
	For the a-spins, we describe their ground states by calculating
	
	\begin{equation}
	\begin{aligned}
	S^{z}_{\mathrm{atot}}=\sum_{\mathrm{ai} \in a}S^{z}_{\mathrm{ai}},
	\end{aligned}
	\end{equation}	
	in which $S^{z}_{\mathrm{ai}}$ stands for the a-spins in the same hexamer. Since $[S^{z}_{\mathrm{atot}},H]=0$, $S^{z}_{\mathrm{atot}}$ is compatible with the Hamiltonian.
		
	\begin{figure}
		\includegraphics[height=7cm]{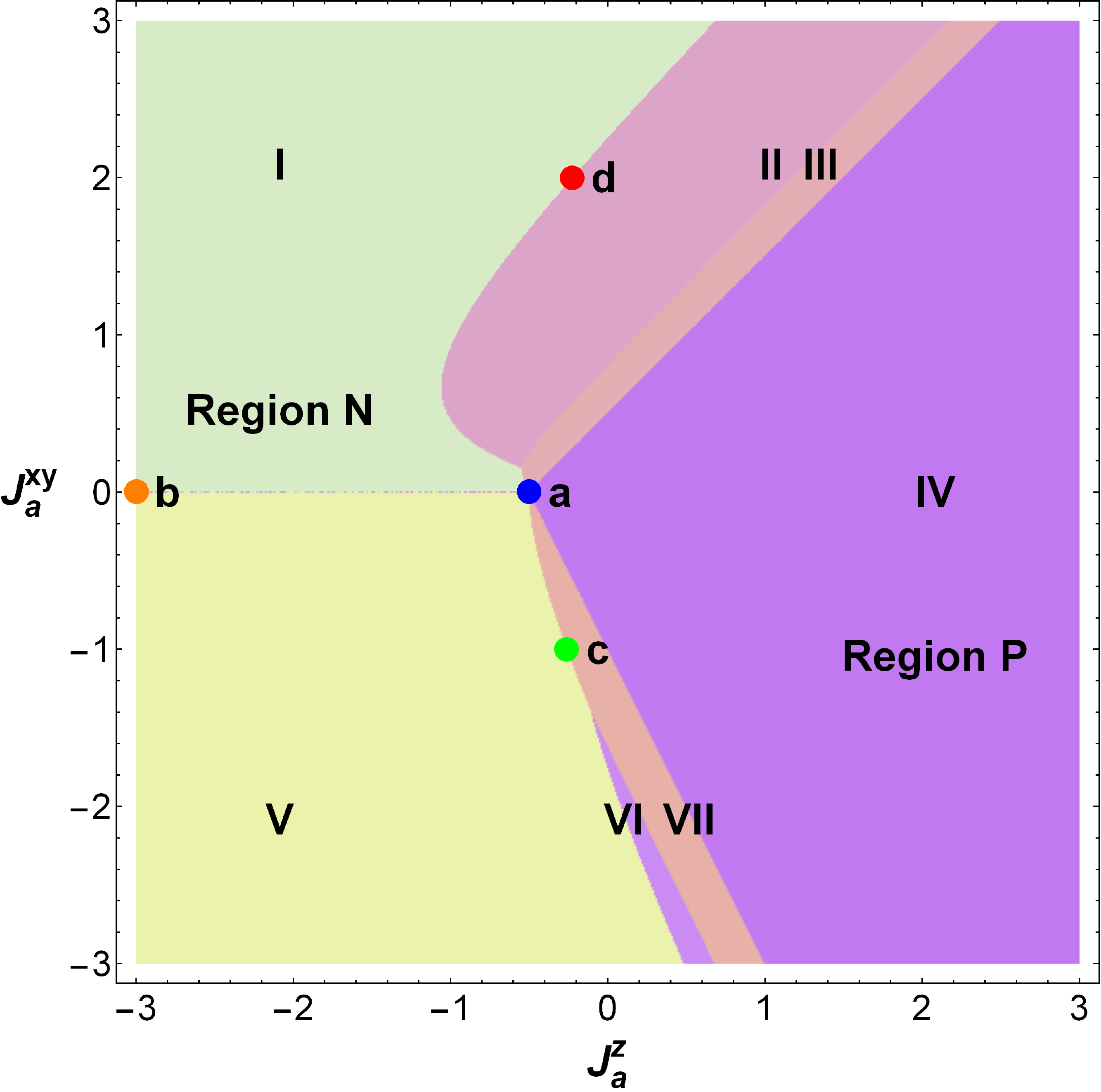}
		\caption{Phase diagram at $T=0$ is identified by the energy of each hexamer when $h=0$ and $J_{ab}^{z}=-1$. The phase diagram is divided into two main regions by the sign of $\beta J_{bb}$, which are denoted as Region P (for $\beta J_{bb}$ is positive) and Region N (for $\beta J_{bb}$ is negative). We mark the phases in Region N as I and V and the phases in Region P as II, III, IV, VII and VI. We also mark some of the highly degenerate points and plot their effective $\beta J_{\mathrm{bb}}$ as functions of temperature}
		\label{Fig4}
	\end{figure}

	\begin{table}[tbp]
		\centering
		\caption{The effective $\beta J_{\mathrm{bb}}$, $S_{\mathrm{atot}}^{z}$ and degeneracy $D$ of each phase for $h=0$ at zero temperature.}
		\begin{tabular}{lcccc}
			\hline
			\hline
			&$\beta J_{\mathrm{bb}}$ &$S_{\mathrm{atot}}^{z}$ &D  &Phase\\
			\hline
			\hline
			Region N &$-\infty$   & & &Antiferromagnetic Phase\\
			\hline
			Region N& Including:&&&\\
			\hline
			Phase I &$-\infty$ &$0$ &1 &Antiferromagnetic Phase\\
			
			Phase V &$-\infty$ &$0$ &2 &Antiferromagnetic Phase\\
			\hline
			\hline
			Region P  &$\infty$    & &  &Ferrimagnetic Phase\\
			\hline
			Region P& Including:&&&\\
			\hline
			Phase II &$\infty$ &$-1$ &1 &Ferrimagnetic Phase\\
			Phase III &$\infty$ &$-2$ &1 &Ferrimagnetic Phase\\
			Phase IV &$\infty$ &$-3$ &1 &Ferrimagnetic Phase\\
			Phase VI &$\infty$ &$-1$ &2 &Ferrimagnetic Phase\\
			Phase VII &$\infty$ &$-2$ &2 &Ferrimagnetic Phase\\
			\hline
			\hline
		\end{tabular}
		\label{Tab1}
	\end{table}	
	\begin{table}[tbp]
		\centering
		\caption{The $\beta J_{\mathrm{bb}}$and entropy per site of the TKL model for phase boundaries when $h=0$ at zero temperature. All of them have the finite correlation length}
		\begin{tabular}{l|c|cc} 		
			\hline
			\hline
			Boundary&Boundary About&$\beta J_{\mathrm{bb}}$ &$s_{0}/9$\\
			\hline
			BL:I &Phase I and Phase V&$\left(1/4\right)\ln\left(3/5\right)$ &$0.407945$\\
			
			BL:II &Phase I and Phase II&$0$ &$0.231049$\\
			
			BL:III &Phase V and Phase VI&$\left(1/4\right)\ln2$ &$0.280644$\\
			
			BL:IV &Phase V and Phase VII&$\left(1/4\right)\ln2$ &$0.280644$\\
			
			BP:I &Phase I, V and IV&$\left(1/4\right)\ln\left(7/5\right)$ &$0.609883$\\
			
			BP:II &Phase V, VI and VII&$\left(1/4\right)\ln4$ &$0.356169$\\
			
			\hline
			\hline
			
		\end{tabular}
		\label{Tab2}
	\end{table}	
	
	By investigating $S^{z}_{\mathrm{atot}}$ and the energy level of each hexamer (Fig.\ref{FigE} for instance), we present the phase diagram at zero temperature in Fig.\ref{Fig4} and Table.\ref{Tab1}. The phase diagram can be divided into two major regions according to the sign of $\beta J_{bb}$, which are Region P for positive and Region N for negative (see Fig.\ref{Fig4}). The boundaries between these regions are denoted as BL:II, BL:III and BL:IV (see Table.\ref{Tab2}) according to the phases on each side of them. Also, Table.\ref{Tab2} gives the entropies of these phase boundaries at zero temperature. Note that the BP:I keeps the highest entropy, which satisfies the intuition that the entropy of a system at transition lines or dots should be higher than that of the surrounding phases.
	
	In Region P, the ground state energies of each hexamer obey $E_0(\uparrow\uparrow\uparrow)<E_0(\downarrow\uparrow\uparrow)$ and $\beta J_{bb}$ tends to infinity. In this case, the b-spins have a perfect ferromagnetic long-range order because $\beta J_{bb}$ exceeds the critical point of the ferromagnetic Kagome Ising model ($\beta J_{\mathrm{kag}}=(\ln[3+\sqrt{12}])/4$)\cite{syozi1951statistics}. Moreover, this region is divided into five phases corresponding to different states of the a-trimers. Their eigenvectors is given in Appendix.\ref{Eigenvector}. Although there is a ferromagnetic order for the b-spins, Region P is in the ferrimagnetic phases since $S^{z}_{\mathrm{atot}}<0$. Lastly, when $J_{a}^{z}$ increases, $S^{z}_{\mathrm{atot}}$ decreases to $-3$ gradually in Region P. 
	
	In Region N, the ground state energies obey  $E_0(\uparrow\uparrow\uparrow)>E_0(\downarrow\uparrow\uparrow)$, meaning that $\beta J_{bb}$ tends to negative infinity. It leads to an antiferromagnetic phase for the b-spins\cite{moessner1999two-dimensional}. This region can also be divided into two phases but both of them correspond to $S_{\mathrm{atot}}^{z}=0$.
	
\begin{figure}
	\includegraphics[height=5cm]{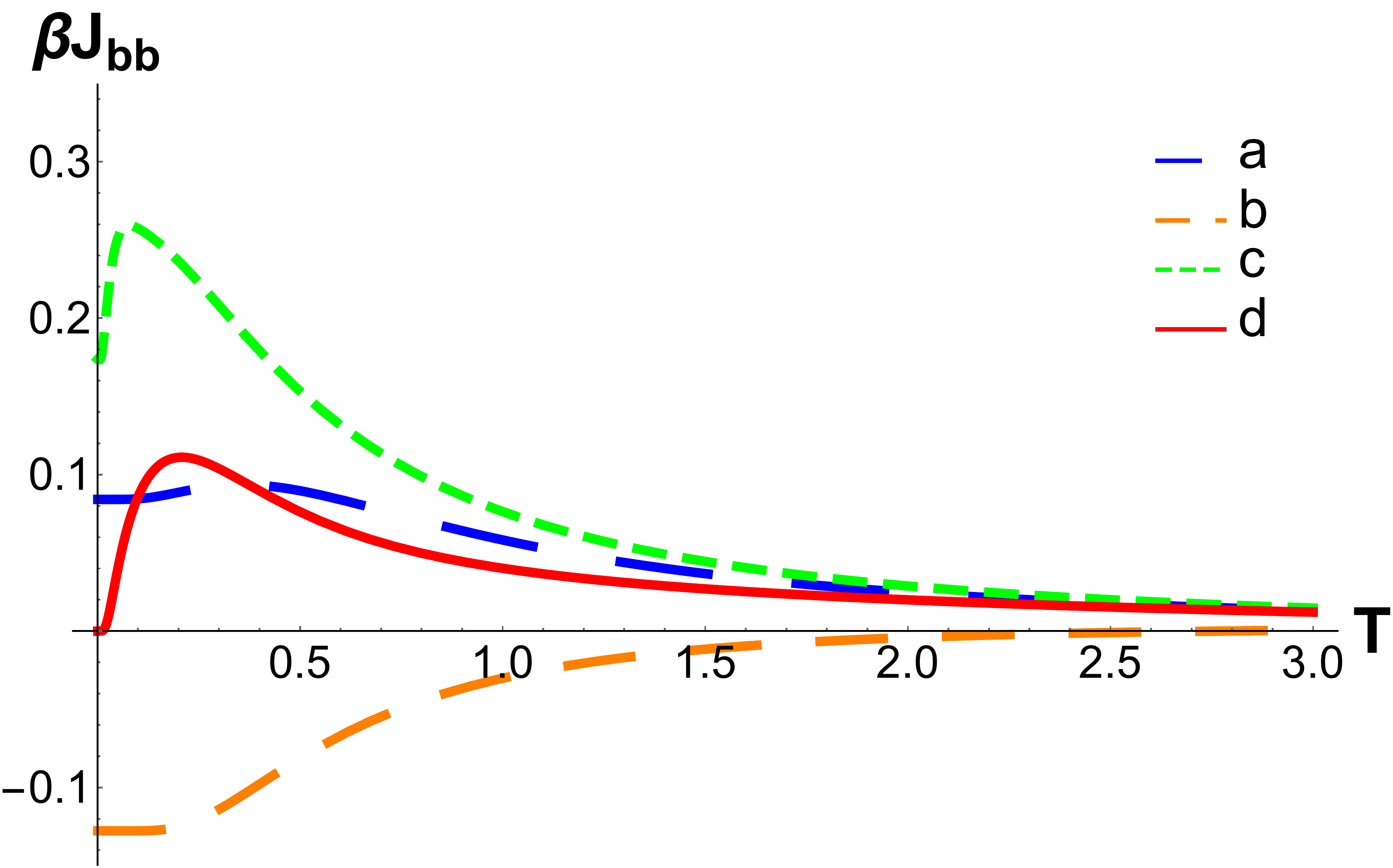}
	\caption{The effective $\beta J_{\mathrm{bb}}$ does not tend to infinity in some of the highly degenerate points at $T=0$. We plot some of their effective $\beta J_{\mathrm{bb}}$ as functions of temperature here. The blue (middle dashed) line is for point a in Fig.\ref{Fig4} with $J_{a}^{z}=-0.5, J_{a}^{xy}=0$, the orange (lower dashed) one is for point b with $J_{a}^{z}=-3, J_{a}^{xy}=0$, the green (upper dashed) one is for point c with $J_{a}^{z}=-0.259402, J_{a}^{xy}=-1$, the red (solid) one is for point d with $J_{a}^{z}=-0.223591, J_{a}^{xy}=2$.}
	\label{FigbetaJ}
	\end{figure}

	At the boundaries between Region P and Region N, the ground state energies are equal,  $E_0(\uparrow\uparrow\uparrow)=E_0(\downarrow\uparrow\uparrow)$. In addition, $E_0(\uparrow\uparrow\uparrow)$ and $E_0(\downarrow\uparrow\uparrow)$ can also be equal in BL:I, BP:I and BP:II (see Table.\ref{Tab2}). In these cases, the value of $\beta J_{bb}$ depends on the ratio of the degeneracies $D_1/D_2$. For different boundaries, the possible values of $\beta J_{bb}$ can be positive (BL:III, BL:IV, BP:I, BP:II) and negative (BL:I) or even zero (BL:II) (see Fig.\ref{FigbetaJ}) at zero temperature, and most of them are not monotonic with temperature. 
	
\begin{figure}[htbp]
		\centering
		{\subfigure[]{
				\includegraphics[height=6.5cm]{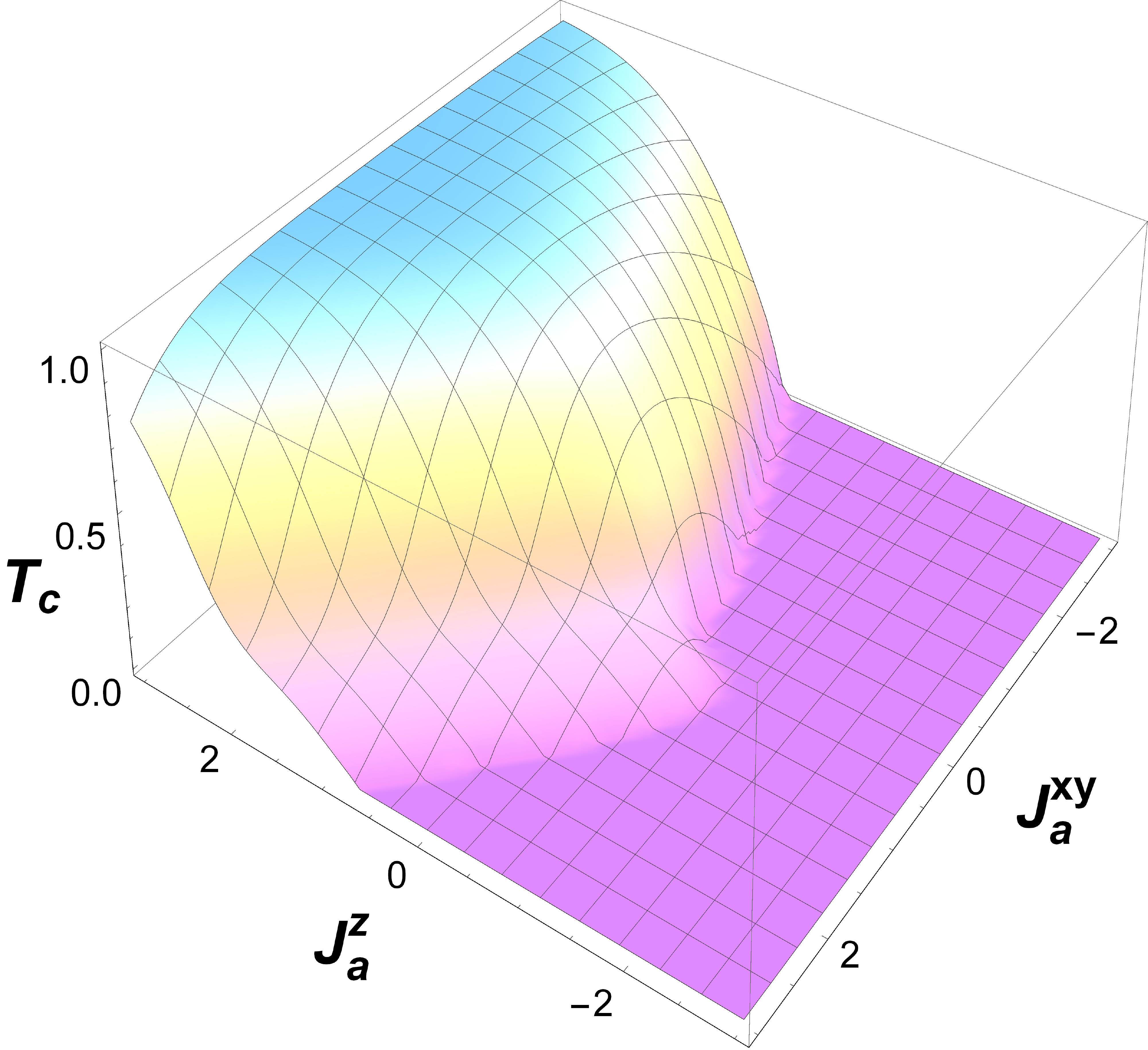}}}
		{\subfigure[]{
				\includegraphics[height=5cm]{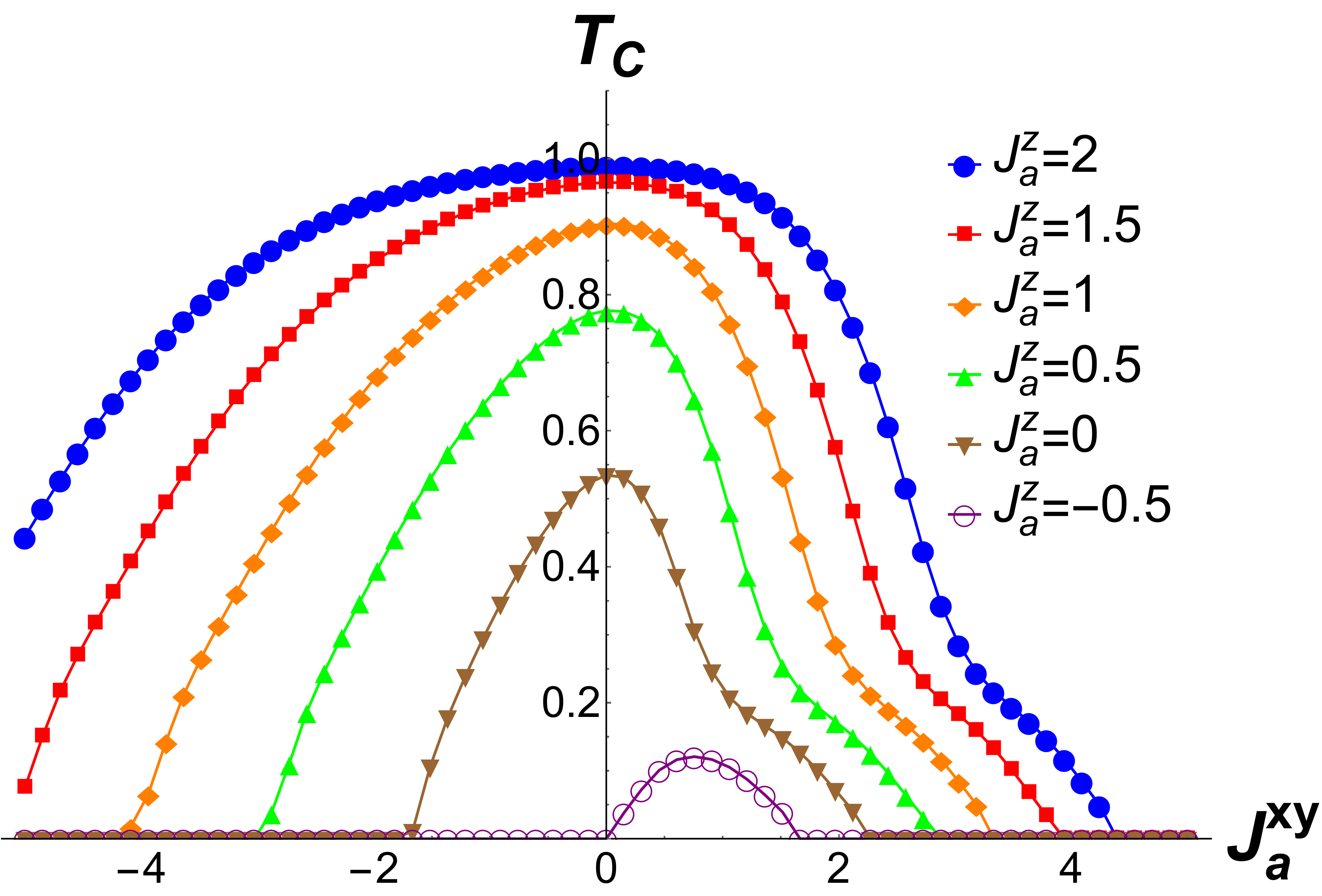}}}
		\caption{Finite-temperature phase diagram of the spin-$1$ TKL model with the $|J_{ab}^{z}|=1$ as the unit of energy. Fig.\ref{Fig_Tc}(a) is the finite temperature phase diagram and Fig.\ref{Fig_Tc}(b) is the critical temperature $T_{C}$ as a function of $J^{xy}_{a}$ when $J^{z}_{a}$ runs from $-0.5$ to $2$.}
		\label{Fig_Tc}
	\end{figure}

	Lastly, Fig.\ref{Fig_Tc} gives the finite-temperature phase diagram as a function of $J^{xy}_{a}$, $J^{z}_{a}$ and critical temperature $T_{C}$ by investigating the critical point of the effective model ($\beta J_{\mathrm{kag}}=(\ln[3+\sqrt{12}])/4$). This fits with the intuition that the disordered phases should become the largest part of the phase diagram when the temperature increases. Moreover, the ferrimagnetic phases with lower $S_{atot}^{z}$ have a higher critical temperature of the spontaneous order. Actually, higher decorated spins in the TKL model can cause a stronger effective coupling between the b-spins.  
	
	\subsection{Physical explanation and comparison with the spin-1/2 TKL model}
	
	To explain how the quantum fluctuations cause these new quantum phases, we consider $J_{a}^{xy} = 0$ first. There is a phase transition point between the disordered phase and the ordered phase, which is $J_{a}^{z}=-0.5$(see Fig.\ref{Fig4}). Each hexamer is non-degenerate in the ordered phase while it is eight-fold degenerate in the disordered phase, which are $D=3$ for the b-spins configuration $(\uparrow\uparrow\uparrow)$ with $S_{\mathrm{atot}}^{z}=-1$ and $D=5$ for $(\downarrow\uparrow\uparrow)$ with $S_{\mathrm{atot}}^{z}=\pm1,0$. However, it is twelve-fold degenerate at the phase transition point, which is higher than the sum of the degeneracies in the disordered phase and the ordered phase. Such a difference comes from an intermediate state ($(\uparrow\uparrow\uparrow)$ with $S_{\mathrm{atot}}^{z}=-2$) at the phase transition point.
	
	When we consider $J_{a}^{xy}$, quantum fluctuations appear and several new phases emerge. $J_{a}^{xy}$ causes this evolution in two different ways. Firstly, it introduces the fluctuation of $S_{\mathrm{atot}}^{z}$. Secondly, it influences the values of $\beta J_{bb}$, which may change the ground state of the b-spins. 
	
	As a result, in the disordered phase, the energy degeneracy between $(\uparrow\uparrow\uparrow)$ and $(\downarrow\uparrow\uparrow)$ vanishes when we consider $J_a^{xy}$. And $(\downarrow\uparrow\uparrow)$ with $S_{\mathrm{atot}}^{z}=0$ is favored. Then it evolves into $(\uparrow\uparrow\uparrow)$ with $S_{\mathrm{atot}}^{z}=-1$ as $J_a^{z}$ increases. Furthermore, at the phase transition point ($J_{a}^{z}=-0.5$, $J_{a}^{xy}=0$), the intermediate state above ($(\uparrow\uparrow\uparrow)$ with $S_{\mathrm{atot}}^{z}=-2$) becomes stable intermediate phase III or phase VII (see Fig.\ref{Fig4}). The appearance of these intermediate phases is due to the decrease of quantum fluctuations and the disentanglement of the a-trimers. Finally, as $J_{a}^{z}$ increases, the spin-$1$ TKL model changes from the antiferromagnetic phase with $S_{\mathrm{atot}}^{z}=0$ to the ferrimagntic phase with $S_{\mathrm{atot}}^{z}=-1$. Then, in the ferrimagntic case, $S_{\mathrm{atot}}^{z}$ of each hexamer decreases from $-1$ to $-3$ step-by-step. This leads to some small magnetization plateaus.
	
	Looking in farther details, the sign of $J_{a}^{xy}$ also makes a difference. We list both the eigenvector and the spin-configuration schematic diagrams of these ferrimagnetic phases in Appendix.\ref{Eigenvector}. When $J_{a}^{xy}$ is positive, the a-trimers stay in the singlet trimerized states in both phase II and phase III. When $J_{a}^{xy}$ is negative, the a-trimers tend to be in a dimerized state. In phase VII, two a-spins become a dimer in each a-trimer. Moreover, in phase VI, it is an anisotropic trimerized state which can be viewed as a two-step dimerizing. In this case, two of the a-spins become a dimer. Then this dimer dimerizes with the last a-spin in each a-trimer. As result, this trimerized state leads to a two-fold degeneracy to each hexamer.
	
	Compared to the pure spin-$1/2$ TKL model, the spin-$1$ TKL model has an antiferromagnetic effective coupling for the b-spins, which makes the geometric frustration of the b-spins play a much more important role and causes a much larger area of the disordered phases in its phase diagram.

	\section{FINITE EXTERNAL FIELD IN THE ZERO TEMPERATURE LIMIT}
	\label{Sec:FFZ}
	
	\subsection{Mapping to the Kagome Ising model with the three-spin coupling}
	
	We now consider the spin-$1$ TKL model with a finite magnetic field, which is parallel to the axis of the b-spins. The transformation method above is also applicable to this case. As a result of the time-reversal symmetry breaking, the odd spin effective coupling terms in the effective Hamiltonian cannot vanish. Finally, it becomes
	\begin{equation}
	\begin{aligned}
	H_{n}'=&-\ln(Z_{a})/\beta-J_{bb}\sum_{\mathrm{bi},\mathrm{bj}\in b}\sigma_{\mathrm{bi}}^z\sigma_{\mathrm{bj}}^z\\&-J_{bbb}\sum_{\mathrm{bi},\mathrm{bj},\mathrm{bk}\in b}\sigma_{\mathrm{bi}}^z\sigma_{\mathrm{bj}}^z\sigma_{\mathrm{bk}}^z-h_{b}\sum_{\mathrm{bi}\in b}\sigma_{\mathrm{bi}}^z,
	\end{aligned}
	\end{equation}
	in which
	\begin{equation}
	\begin{aligned}
	Z_{a}=[Z(\uparrow\uparrow\uparrow,h)Z(\downarrow\downarrow\downarrow,h)]^{\frac{1}{8}}[Z(\downarrow\uparrow\uparrow,h)Z(\uparrow\downarrow\downarrow,h)]^{\frac{3}{8}},
	\end{aligned}
	\end{equation}
	\begin{equation}
	\begin{aligned}
	\beta J_{bb}=\frac{1}{8}\ln\left[\frac{Z(\uparrow\uparrow\uparrow,h)Z(\downarrow\downarrow\downarrow,h)}{Z(\downarrow\uparrow\uparrow,h)Z(\uparrow\downarrow\downarrow,h)}\right],
	\end{aligned}
	\end{equation}
	\begin{equation}
	\begin{aligned}
	\beta J_{bbb}=\frac{1}{8}\ln\left[\frac{Z(\uparrow\uparrow\uparrow,h)Z(\uparrow\downarrow\downarrow,h)^3}{Z(\downarrow\downarrow\downarrow,h)Z(\downarrow\uparrow\uparrow,h)^3}\right],
	\end{aligned}
	\end{equation}
	\begin{equation}
	\begin{aligned}
	\beta h_{b}=\frac{1}{8}\ln\left[\frac{Z(\uparrow\uparrow\uparrow,h)Z(\downarrow\uparrow\uparrow,h)}{Z(\downarrow\downarrow\downarrow,h)Z(\uparrow\downarrow\downarrow,h)}\right].
	\end{aligned}
	\end{equation}
	
	In this case, the new parameters in the effective model ($Z_{a}$, $h_{b}$, $J_{bb}$ and $J_{bbb}$) are the functions of the original couplings in the spin-$1$ TKL model ($J_{ab}^{z}$, $J_{a}^z$, $J_{a}^{xy}$ and $h$).
	
	\subsection{Phase diagram at zero temperature and ground states properties}
	
	Although the odd spin couplings make it hard to obtain a rigorous solution of the model, it is still possible to deduce a full phase diagram of the spin-$1$ TKL model at zero temperature since its effective model is classical\cite{yao2008xxz,moessner1999two-dimensional}. The phase diagram can be achieved in the usual manner by searching for which b-spins configurations of each hexamer keeps the lowest energy. These energies can be written as
	
	\begin{equation}
	\begin{aligned}
	E(\uparrow\uparrow\uparrow)&=-\ln(Z_{a})/\beta-3J_{bb}-J_{bbb}-3h_{b},
	\end{aligned}
	\end{equation}	
	\begin{equation}
	\begin{aligned}
	E(\downarrow\downarrow\downarrow)&=-\ln(Z_{a})/\beta-3J_{bb}+J_{bbb}+3h_{b},
	\end{aligned}
	\end{equation}	
	\begin{equation}
	\begin{aligned}
	E(\downarrow\uparrow\uparrow)&=-\ln(Z_{a})/\beta+J_{bb}+J_{bbb}-h_{b},
	\end{aligned}
	\end{equation}
	\begin{equation}
	\begin{aligned}
	E(\uparrow\downarrow\downarrow)&=-\ln(Z_{a})/\beta+J_{bb}-J_{bbb}+h_{b}.
	\end{aligned}
	\end{equation}
		       	
	 Here, we obtain the ground state of the spin-$1$ TKL model by finding the ground state of each hexamer numerically for each combination of parameters. When the ground state energy of the  b-spin configuration $(\uparrow\uparrow\uparrow)$ or $(\downarrow\downarrow\downarrow)$ is favored, the spin-$1$ TKL model is in ferromagnetic phase or ferrimagnetic phase, which depends on the a-trimer states. For the $(\downarrow\uparrow\uparrow)$ or $(\uparrow\downarrow\downarrow)$ case, the macroscopic ground state of its effective model can be achieved by enumerating the ways of tilling the corresponding effective trimers in the Kagome plane, which is equivalent to placing dimers on the bonds of a honeycomb lattice\cite{yao2008xxz,moessner1999two-dimensional}.
	
	\begin{table}[tbp]
		\centering
		\caption{The energy of the hexamers at some points in each phases with $J_{a}^z =J_{a}^{xy}$, $J_{ab}^{z}=-1$ and different $h$ in Fig.\ref{Fig6}(b). We also highlight (the underlined numbers) which configuration of b-spins has the lowest energy in each phase at zero temperature.}
		\begin{tabular}{l|cc|cccc} 		
			\hline
			\hline
			Phase &$J_{a}^{xy}=J_{a}^{z}$&$h$&$E(\uparrow\uparrow\uparrow)$ &$E(\downarrow\uparrow\uparrow)$ &$E(\downarrow\downarrow\downarrow)$ &$E(\uparrow\downarrow\downarrow)$\\
			\hline
			\hline
			Phase I&-3&1&\underline{-9.75} &-9.46005 &-8.25 &-8.96005\\
			Phase II &-2&3.552&\underline{-9.216} &-9.03515 &-6.44 &-7.664\\
			Phase III &-1.184&3.856&\underline{-8.604} &-8.39904 &-8.124 &-8.052\\
			Phase IV &5&5&\underline{-30.75} &-30.25 &-29.25 &-29.75\\
			Phase V &-1.2&3.4&-7.35 &\underline{-7.37237} &-7.05 &-6.95\\
			Phase VI &2&2&-10.5 &-11.5 &\underline{-13.5} &-12.5\\
			Phase VII &-1&1.2&-3.9 &\underline{-4.16246} &-3.5 &-3.9\\
			Phase VIII &-1&0.3&-3.225 &\underline{-3.55619} &-3.075 &-3.40619\\
			Phase IX &-0.64&0.544&-2.328 &-2.67641 &-2.68 &\underline{-2.688}\\
			Phase X &-0.48&0.32&-1.88 &-2.19117 &\underline{-2.4} &-2.2\\
			
			\hline
			\hline
			
		\end{tabular}
		\label{TabE}
	\end{table}
	
	\begin{figure}[t]
		\centering
		{
			\subfigure[]{
			\includegraphics[height=6.5cm]{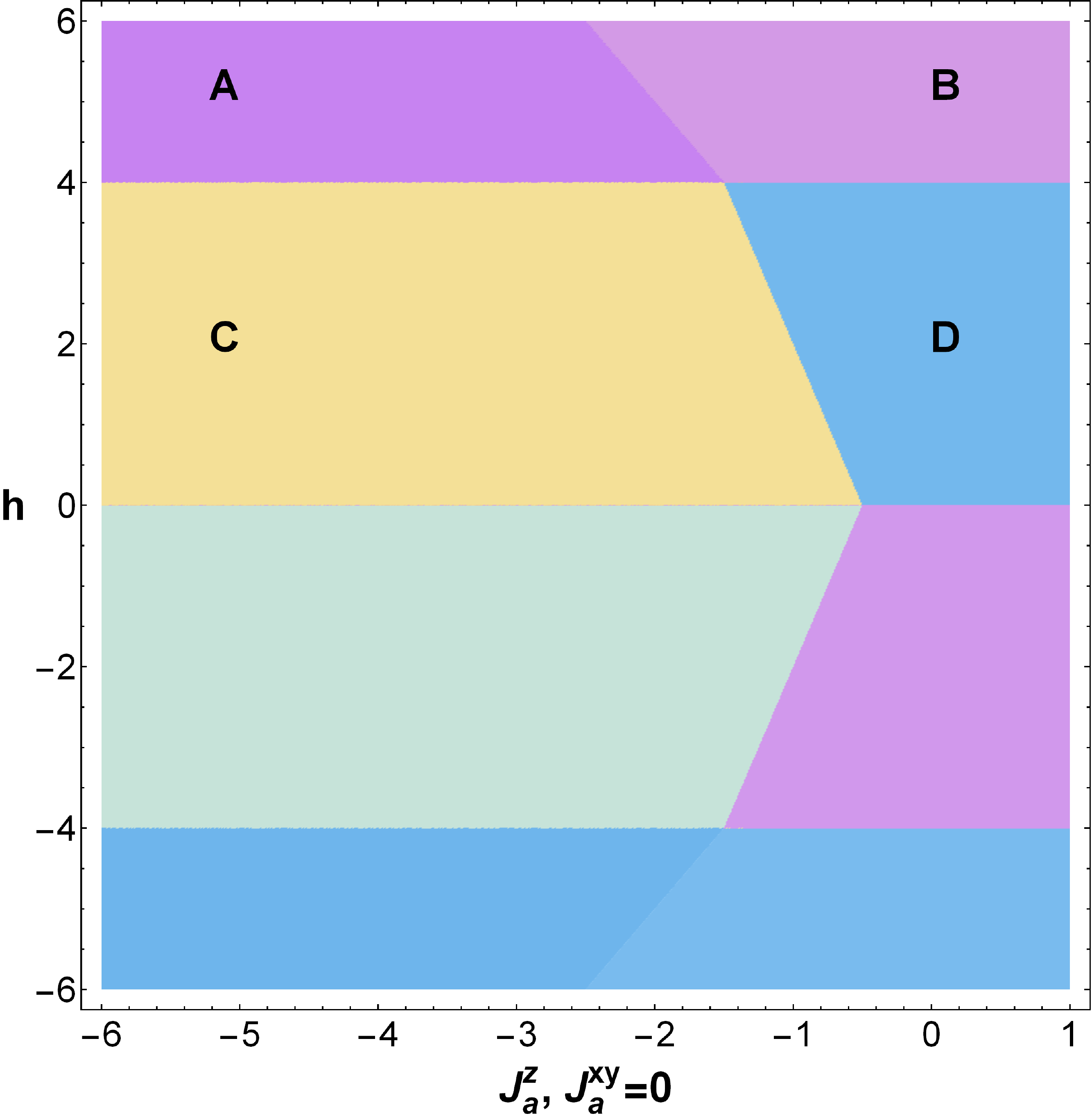}}
		}
		{
			\subfigure[]{
			\includegraphics[height=6.5cm]{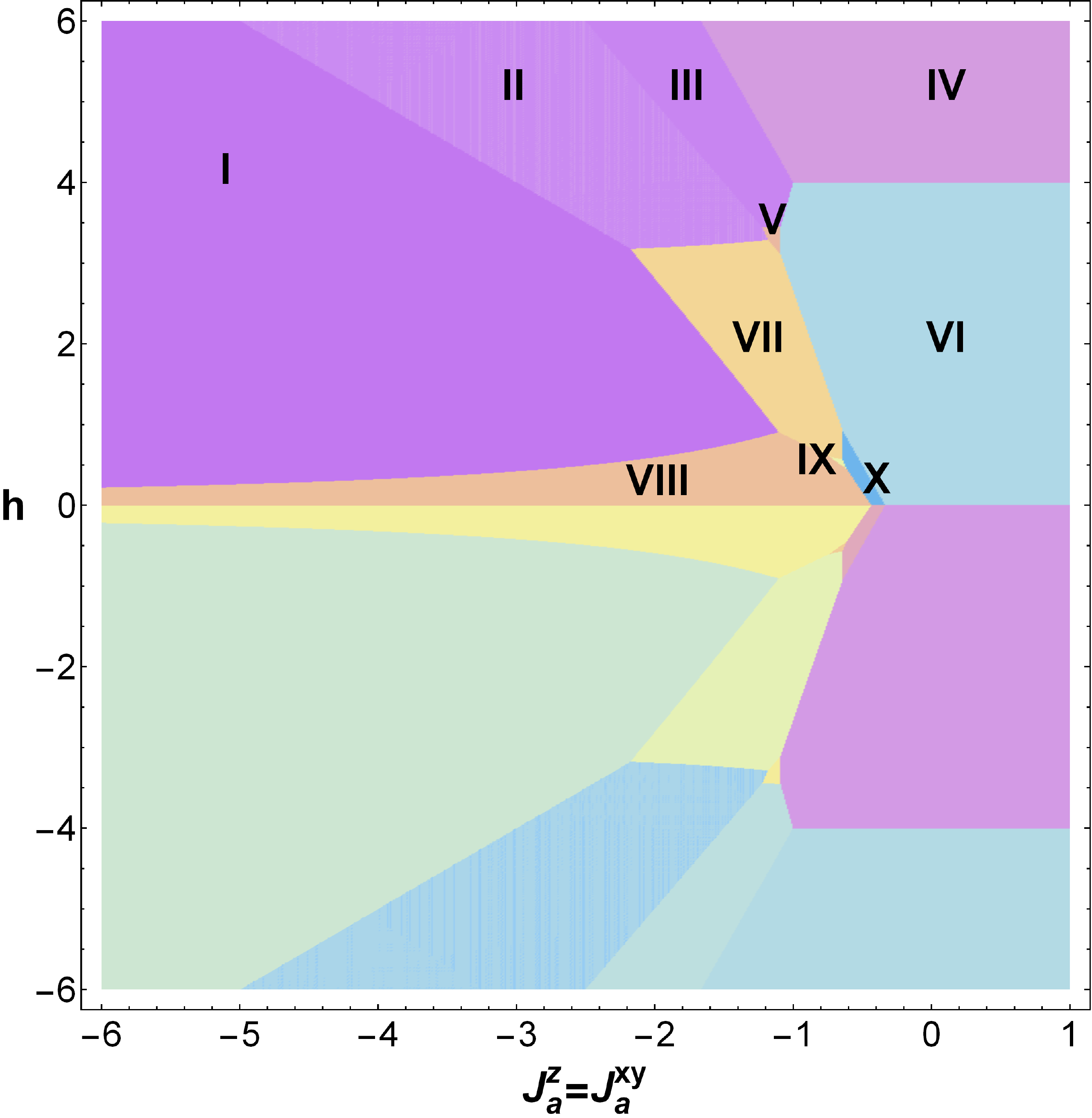}}
		}
		\caption{The phase diagram of the TKL model with spin-$1$ decorated trimers at zero temperature identified by calculating the energy of each hexamer when the spin-$1$ TKL model is in its classical limit ($J_{\mathrm{a}}^{xy}=0$, $J_{\mathrm{ab}}^{z}=-1$ and $T=0$) and Heisenberg limit ($J_{\mathrm{a}}^z =J_{\mathrm{a}}^{xy}$, $J_{\mathrm{ab}}^{z}=-1$ and $T=0$).}
		\label{Fig6}
	\end{figure}
	
	\begin{table}[tbp]
		\centering
		\caption{Each phase in Fig.\ref{Fig6} identified by calculating the configuration of b-spins and $S_{\mathrm{atot}}^{z}$ in the ground state of each hexamer.}
		\label{Tab3}
		\begin{tabular}{ccccc}
			\hline
			\hline
			 Phase Mark&b-spins &$S_{\mathrm{atot}}^{z}$ &D &Phase\\
			\hline
			\hline
			&$J_{\mathrm{a}}^{xy}=0$&$J_{\mathrm{ab}}^{z}=-1$\\
			\hline
			\hline
			Phase A &$\uparrow\uparrow\uparrow$   &$1$   &$3$ &Ferrimagnetic\\
			Phase B  &$\uparrow\uparrow\uparrow$    &$3$     &$1$ &Saturated Ferromagnetic\\
			Phase C  &$\downarrow\uparrow\uparrow$    &$1$     &$1$ &Honeycomb Dimer Liquid\\
			Phase D  &$\downarrow\downarrow\downarrow$    &$3$     &$1$ &Ferrimagnetic\\
			\hline
			\hline
			&$J_{\mathrm{a}}^z =J_{\mathrm{a}}^{xy}$&$J_{\mathrm{ab}}^{z}=-1$\\
			\hline
			\hline
			Phase I &$\uparrow\uparrow\uparrow$   &$0$   &$1$ &Ferrimagnetic\\
			Phase II  &$\uparrow\uparrow\uparrow$    &$1$     &$3$ &Ferrimagnetic\\
			Phase III  &$\uparrow\uparrow\uparrow$    &$2$     &$2$ &Ferrimagnetic\\
			Phase IV  &$\uparrow\uparrow\uparrow$    &$3$     &$1$ &Saturated Ferromagnetic\\
			Phase V &$\downarrow\uparrow\uparrow$   &$2$   &$2$ &Honeycomb Dimer Liquid\\
			Phase VI  &$\downarrow\downarrow\downarrow$    &$3$     &$1$ &Ferrimagnetic\\
			Phase VII  &$\downarrow\uparrow\uparrow$    &$1$     &$1$ &Honeycomb Dimer Liquid\\
			Phase VIII  &$\downarrow\uparrow\uparrow$    &$0$     &$1$ &Honeycomb Dimer Liquid\\
			Phase IX &$\downarrow\downarrow\uparrow$   &$1$   &$1$ &Honeycomb Dimer Liquid\\
			Phase X  &$\downarrow\downarrow\downarrow$    &$2$     &$2$ &Ferrimagnetic\\			
			\hline
			\hline
		\end{tabular}
	\label{TabPhase}
	\end{table}
	
	Fig.\ref{Fig6} is the phase diagram when $J_{a}^{xy}=0$, $J_{ab}^{z}=-1$ (Fig.\ref{Fig6}(a)) and $J_{a}^z =J_{a}^{xy}$, $J_{ab}^{z}=-1$ (Fig.\ref{Fig6}(b)) at zero temperature. Table.\ref{TabE} lists the ground state energy of each hexamer at selected points in each phase of Fig.\ref{Fig6}(a) and Fig.\ref{Fig6}(b). The phase diagram is divided into eight parts when $J_{a}^{xy}=0$ but twenty in the $J_{a}^z =J_{a}^{xy}$ case. Both of them are symmetric about $h=0$.
	
	The phase diagram Fig.\ref{Fig6}(a) is similar to its counterpart in the pure spin-$1/2$ TKL model. However, Fig.\ref{Fig6}(b) case is quite different, including the absence of the Kagome loop gas phase\cite{yao2008xxz} and the presence of some unstable phases (phase V, phase IX, and phase X).
	
	\subsection{Physical explanation and the effect of the a-trimer quantum fluctuations}
	
	To explain how these unstable phases come from the quantum fluctuations of the a-trimers, we start with the classical limit ($J_{a}^{xy}=0$). In Fig.\ref{Fig6}(a), the phase diagram is divided into four different phases when $h$ is positive, including saturated ferromagnetic phase (Phase B), ferromagnetic phase with $S_{\mathrm{atot}}^{z}=1$ (Phase A), ferrimagnetic phase with $S_{\mathrm{atot}}^{z}=3$ (Phase D) and honeycomb dimer phase (Phase C). It is quite similar to the spin-$1/2$ TKL case\cite{yao2008xxz}. Their boundary conditions can be given by $2J_{a}^{z}+h=1$ for the boundary between Phase A and Phase B; $h=4$ for the boundary between Phase A and Phase C or between Phase B and Phase D; $h+4J_{a}^{z}=-2$ for the boundary between Phase C and Phase D.
	
	\begin{figure}
		\centering
		{\subfigure[]{
			\includegraphics[height=6.5cm]{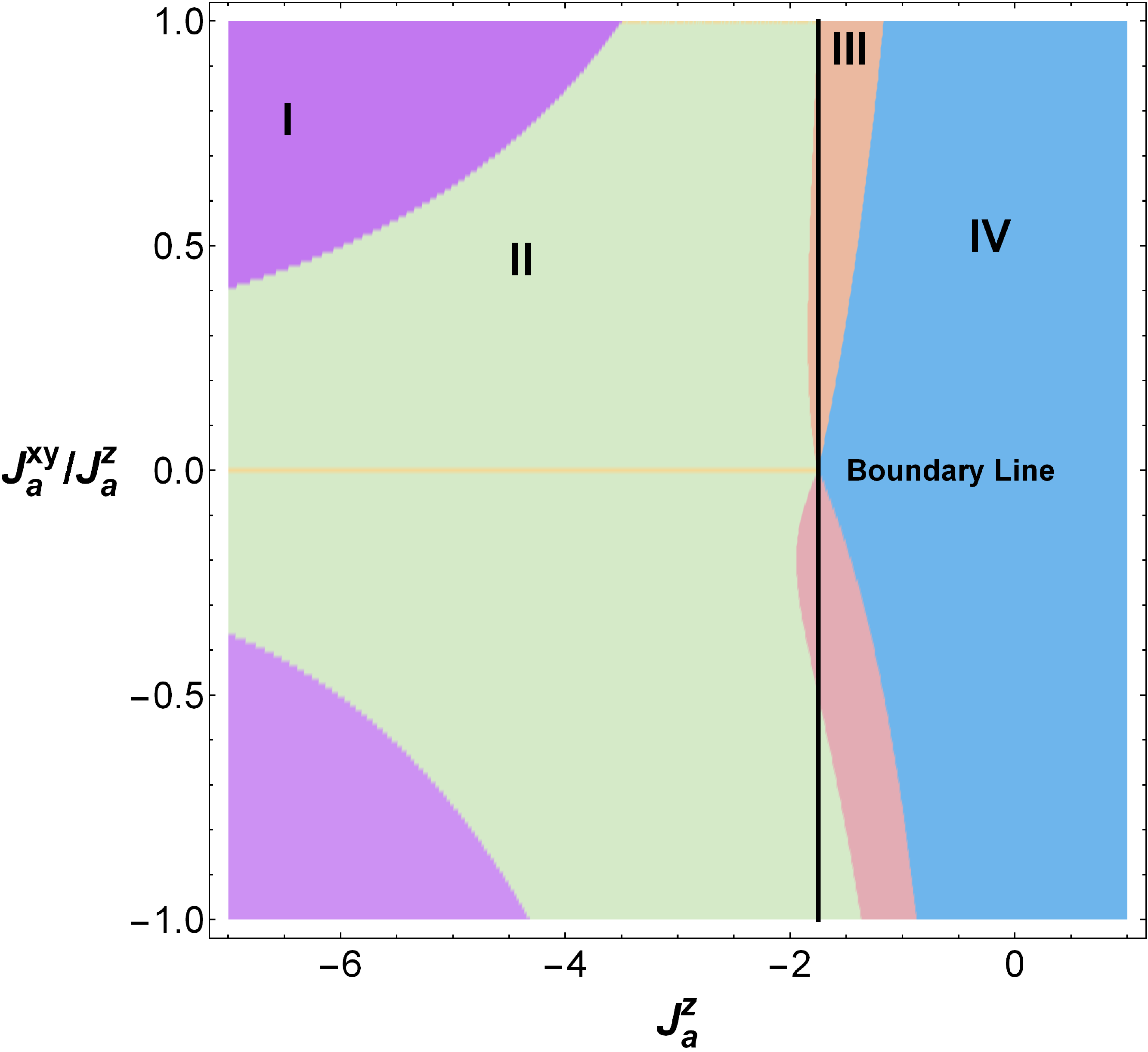}}
		\label{Fig7a}}
		{\subfigure[]{
			\includegraphics[height=6.5cm]{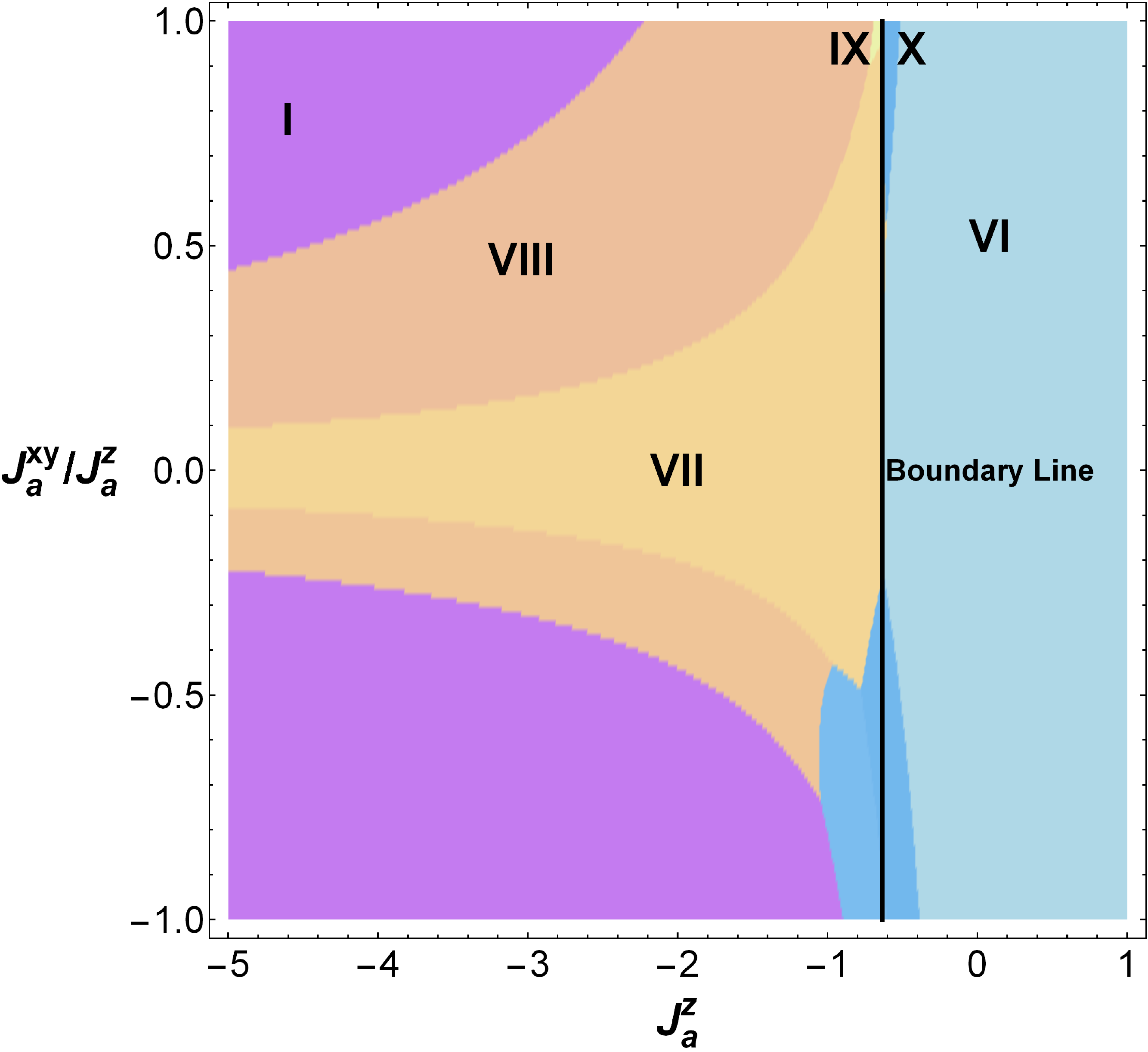}}
		\label{Fig7b}}
		\caption{These figures present the phases changing as  $J_{a}^{xy}/J_{a}^{z}$ increases from $0$ to $1$. These phase marks here are same as in Fig.\ref{Fig6}(b). $J_{ad}^{z}=-1$ and $h=4.5$ in Fig.\ref{Fi7}(a) while $J_{ad}^{z}=-1$ and $h=0.544$ in Fig.\ref{Fi7}(b), which is crossing through the unstable phase IX. The boundary line in the figures is crossing through the transition point in the classical limit with the same values of $J_{ad}^{z}$. These figures present a better view of how new phases emerge.}
		\label{Fi7}  		
	\end{figure}
	
	Fig.\ref{Fi7} tells us how the phase diagram evolves with the localized quantum fluctuations of the a-trimers, in which $h$ and $J_{ab}^{z}$ are constants, and $J_{a}^{xy}/J_{a}^{z}$ and $J_{a}^{z}$ serve as parameters. We present these phase diagrams with $J_{ad}^{z}=-1$ and $h=4.5$ in Fig.\ref{Fi7}(a) and $J_{ad}^{z}=-1$ and $h=0.544$ in Fig.\ref{Fi7}(b). The former contains ordinary stable phases while the latter includes the most unstable phase (phase IX) in its Heisenberg limit ($J_{a}^z =J_{a}^{xy}$).
	
	In Fig.\ref{Fi7}(a), the spin-$1$ TKL model changes into phase IV (see Table.\ref{TabPhase}) directly in its classical limit as $J_{a}^{z}$ increases. As a result, there is a phase transition at $J_{a}^{z}=-1.75$ and $J_{a}^{xy}/J_{a}^{z}=0$ in Fig.\ref{Fi7}(a). When $J_{a}^{xy}/J_{a}^{z}$ increases, Phase III emerges from its phase transition point. If we set a boundary line crossing this phase transition point, we can see that phase III evolves along with this boundary line (see Fig.\ref{Fi7}(a)). Meanwhile, phase I emerges from the left side of the phase diagram. Eventually, in its Heisenberg limit, the spin-$1$ TKL model changes from phase I, to phase II, then to phase III, and finally to phase IV. This phase transition corresponds to a gradual disentanglement of the a-trimers microscopically and leads to a stable growth of magnetization plateaus macroscopically as $h$ increases (see Fig.\ref{Fig11})\cite{cisarova2013intermediate}, which is shown in Appendix.\ref{Eigenvector}.  
	
	\begin{figure}
		\centering
		{\includegraphics[height=3.5cm]{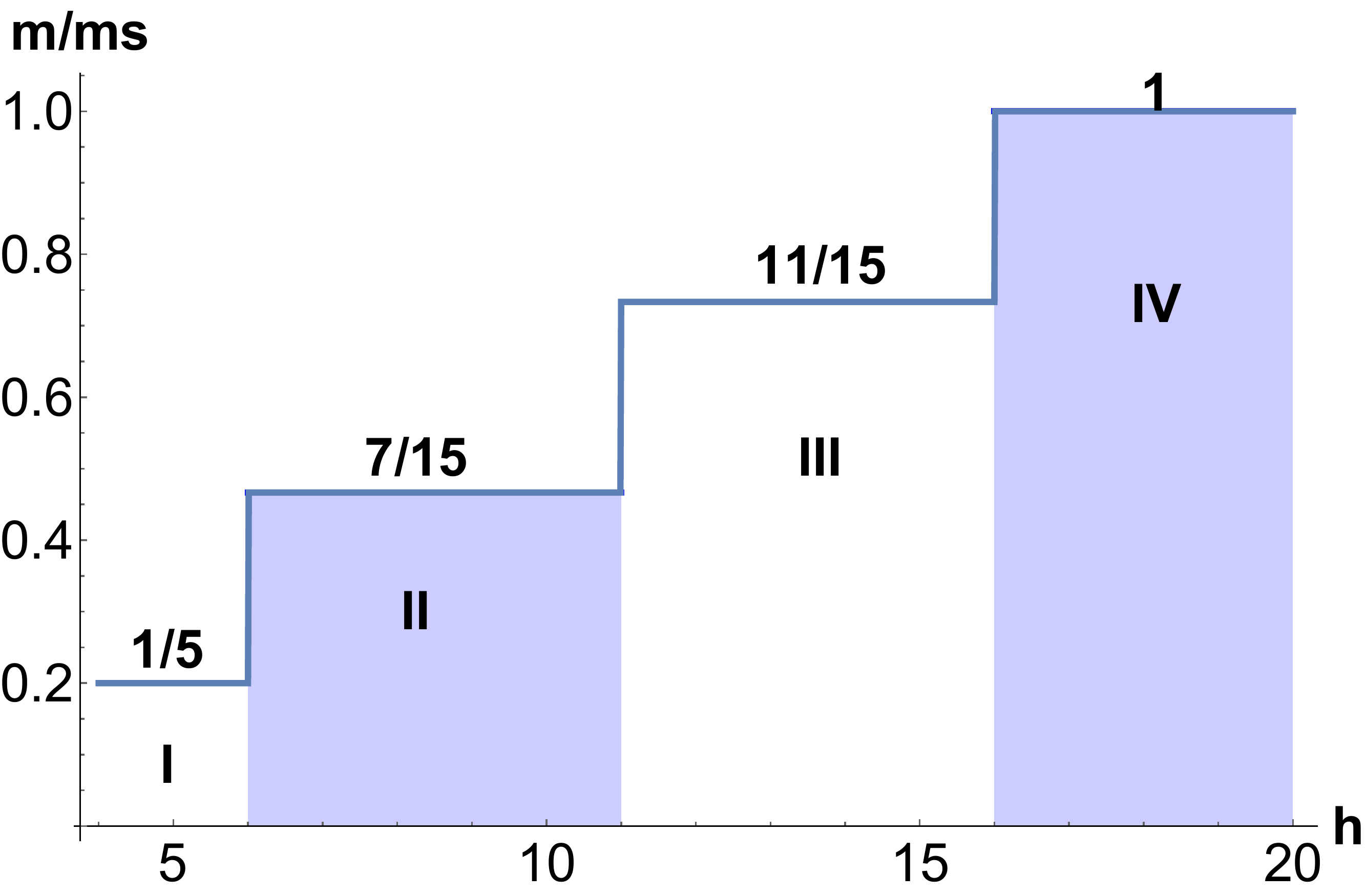}}
		\caption{This figure shows the stably growth of magnetization plateaus with $h$ running from 4 to 20 when $J_{a}^{z}=J_{a}^{xy}=-5$. These phases marked here are the same with those in Fig.\ref{Fig6}(b).}
		\label{Fig11}  		
	\end{figure}
	
	Unlike the Fig.\ref{Fi7}(a) case, $J_{a}^{z}$ already makes the b-spins ground state change from $(\downarrow\uparrow\uparrow)$ to $(\downarrow\downarrow\downarrow)$ in the classical limit (see Fig.\ref{Fi7}(b)). We also set a boundary line crossing through the phase transition point in Fig.\ref{Fi7}(b). Around this boundary line, the b-spins ground state tends to change. Firstly, when $J_{a}^{xy}$ increases, it causes the a-trimers to evolve independently, which makes phase VIII and phase X emerge. Secondly, as $J_{a}^{xy}$ increases, phase VIII and phase X come close to the boundary line. When they are close enough to each other, phase IX emerges around the boundary line due to the unstable b-spins ground state around the boundary line. phase IX can be viewed as an intermediate phase of the phase transition from phase VIII to phase X. Finally, in the Heisenberg limit, the b-spins configuration of the hexamers changes from $(\downarrow\uparrow\uparrow)$, to $(\downarrow\downarrow\uparrow)$, and lastly to $(\downarrow\downarrow\downarrow)$  as $J_{a}^{z}$ increases. Similarly, those unstable phases (phase V, phase IX, and phase X) are the intermediate phases in Fig.\ref{Fig6}(b).
	
	\subsection{Comparison with the spin-1/2 TKL model and discussion of the higher spin TKL model}
	
	We now compare the phase diagram of the spin-$1$ TKL model with the spin-$1/2$ case and find that the spin-$1$ TKL model has a much more diversified phase diagram at zero temperature. These differences mainly come from the richer possible states of the a-trimers under the interplay between its geometric frustration and quantum fluctuations.
	
	In the classical limit, there are four main phases both in the spin-$1/2$ and in the spin-$1$ cases, including the Saturated Ferrimagnetic phase $(\uparrow\uparrow\uparrow)$, the Honeycomb Dimer Liquid phase $(\downarrow\uparrow\uparrow)$, the Ferrimagnetic phase $(\uparrow\uparrow\uparrow)$ and the Ferrimagnetic phase $(\downarrow\downarrow\downarrow)$. When $J_{a}^{xy}$ appears, new phases emerge from the boundary between these main phases. Lastly, the spin-$1$ TKL model has stable magnetization plateaus in the Heisenberg limit. Such effect can also be obtained in the Heisenberg-Ising diamond chain\cite{takano1996ground,canova2009exact}.
	
	Although it is hard to give an exact picture of how the phase diagram of the TKL model develops when the spins in the a-trimers tend to infinity, it is still possible to give its simple description. Firstly, the four main phases above still exist in their classical limit. These main phases also remain in their Heisenberg limit. Besides, more new phases with different a-trimer states emerge from the phase boundaries between these four main phases when we consider higher spins in the a-trimers. This is due to the larger Hilbert space of the decorated trimers. Moreover, when considering higher spins in the a-trimers, some of these new phases may appear more compactly around these phase boundaries. Eventually, if the decorated spins tend to infinity, they would tend to become a line or dot with high degeneracy and the phase diagram in the Heisenberg limit at zero temperature should approach its classical limit.
	
	\section{CONCLUSION}
	\label{Sec:CONCLUSION}
	In conclusion, we have discussed the TKL XXZ-Ising model decorated by the spin-$1$ a-trimers and computed its phase diagram by transforming into an effective Kagome Ising model with or without the three-spin coupling according to the presence of the finite external field. The transformation method is an essentially algebraic method and can be applied even in more general cases.
	
	For $h=0$, the effective model can be simplified with the $C3$ symmetry and the time-reversal symmetry. And the spin-$1$ TKL model is mapped to the Kagome Ising model exactly. In the zero temperature phase diagram, there are two main regions corresponding to the ferromagnetic effective coupling and antiferromagnetic case respectively. Each main region is divided into several phases corresponding to different entangled states of the a-trimers. Compared to the spin-$1/2$ TKL model, one of the most interesting differences of the spin-$1$ TKL model is that the spin-$1$ decorated trimers introduce the antiferromagnetic effective coupling of the b-spins to the TKL model.
	
	When there is a finite external field, only the $C3$ symmetry can be applied to simplify the effective model, which means that we can map the spin-$1$ TKL model to the Kagome Ising model with the three-spin coupling. We give its phase diagram at zero temperature since its effective model is classical\cite{yao2008xxz,moessner1999two-dimensional,ishikawa2014kagome-triangular}. For the classical limit ($J_{a}^{xy}=0$), there are four main phases, which is similar to the spin-$1/2$ TKL case. However, in the presence of $J_{a}^{xy}$, more new phases emerge in the spin-$1$ TKL model than in the spin-$1/2$ case. Finally, in the Heisenberg limit ($J_{a}^{xy}=J_{a}^{z}$),  the spin-$1$ TKL model has several stable magnetization plateaus which correspond to the gradual disentanglement of the a-trimers.
	
	The higher spins in the decorated a-trimers can give a larger dimension of Hilbert spaces. In the spin-$1/2$ case, the possible values of $S_{\mathrm{atot}}^{z}$ are $\pm1/2$ and $\pm3/2$ in both the classical limit and the Heisenberg limit. However, in the spin-$1$ case, the possible values of $S_{\mathrm{atot}}^{z}$ are $\pm1$and $\pm3$ in the classical limit but $0$, $\pm1$, $\pm2$ and $\pm3$ in the Heisenberg limit. This leads to more plentiful states for the a-trimers and more complicated effective couplings. Therefore, the phase diagram becomes more diversified for the spin-$1$ TKL model compared with the spin-$1/2$ case. This is a strong evidence that the quantum fluctuations can create new phases in the highly frustrated spin systems, and help us to understand how the XXZ-Ising decorated model evolves into its classical limit when $S$ increases for the decorated spins.
	
	\appendix
	\section{}\label{Explicit Form}
	Here, we give the explicit formulas of $Z(\uparrow\uparrow\uparrow,h)$, $Z(\downarrow\downarrow\downarrow,h)$, $Z(\downarrow\uparrow\uparrow,h)$ and $Z(\uparrow\downarrow\downarrow,h)$. We use the time reversal symmetry to simplify our description, which is given by $Z(\uparrow\uparrow\uparrow,h)=Z(\downarrow\downarrow\downarrow,-h)$ and $Z(\downarrow\uparrow\uparrow,h)=Z(\uparrow\downarrow\downarrow,-h)$.
	\begin{equation}
	\begin{aligned}
	&Z(\uparrow\uparrow\uparrow,h)=2e^{-\frac{1}{4}\beta\left(-4J_a^{xy}+4 J_a^{z}-3 h\right)}+2e^{-\frac{1}{4}\beta\left(4J_a^{xy}+4J_a^{z}-3h\right)}\\&+e^{-\frac{1}{4}\beta\left(8 J_a^{xy}+4J_a^{z}-3h\right)}+2e^{-\frac{1}{4}\beta\left(4 J_a^{xy}-4 J_a^{z}+8 J_{ab}^{z}+5 h\right)}\\&+2e^{-\frac{1}{4}\beta\left(4 J_a^{xy}-4 J_a^{z}-8 J_{ab}^{z}-11 h\right)}+e^{\frac{3}{4}\beta \left(4 J_a^{z}-4 J_{ab}^{z}-3 h\right)}\\&+e^{-\frac{1}{4}\beta\left(-8J_a^{xy}-4 J_a^{z}+8 J_{ab}^{z}+5h\right)}+e^{-\frac{1}{4}\beta\left(-8 J_a^{xy}-4J_a^{z}-8 J_{ab}^{z}-11 h\right)}\\&+e^{\frac{3}{4}\beta\left(4 J_a^{z}+4 J_{ab}^{z}+5h\right)}\\&+2e^{-\frac{1}{4}\beta\left(2J_a^{xy}+2J_a^{z}+4 J_{ab}^{z}+h-2\sqrt{5 (J_a^{xy})^2+(J_a^{z})^2-2 J_a^{xy}J_a^{z}}\right)}\\&+2e^{-\frac{1}{4} \beta  \left(2J_a^{xy}+2J_a^{z}+4J_{ab}^{z}+h+2\sqrt{5 (J_a^{xy})^2+(J_a^{z})^2-2J_a^{xy}J_a^{z}}\right)}\\&+2 e^{-\frac{1}{4}\beta\left(2 J_a^{xy}+2 J_a^{z}-4J_{ab}^{z}-7h-2\sqrt{5(J_a^{xy})^2+(J_a^{z})^2-2J_a^{xy}J_a^{z}}\right)}\\&+2e^{-\frac{1}{4}\beta\left(2J_a^{xy}+2J_a^{z}-4 J_{ab}^{z}-7h+2 \sqrt{5(J_a^{xy})^2+(J_a^{z})^2-2 J_a^{xy}J_a^{z}}\right)}\\&+e^{-\frac{1}{4}\beta\left(-4J_a^{xy}+2 J_a^{z}-3h-2\sqrt{28(J_a^{xy})^2+(J_a^{ z})^2-4J_a^{xy}J_a^{z}}\right)}\\&+e^{-\frac{1}{4}\beta\left(-4J_a^{xy}+2 J_a^{z}-3h+2\sqrt{28(J_a^{xy})^2+(J_a^{z})^2-4J_a^{xy}J_a^{z}}\right)}\\&+e^{-\frac{1}{4}\beta\left(-4J_a^{xy}+2 J_a^{z}-4 J_{ab}^{z}-7h-2 \sqrt{20(J_a^{xy})^2+(J_a^{z})^2+4 J_a^{xy}J_a^{z}}\right)}\\&+e^{-\frac{1}{4}\beta\left(-4J_a^{xy}+2J_a^{z}-4J_{ab}^{z}-7h+2\sqrt{20(J_a^{xy})^2+(J_a^{z})^2+4J_a^{xy}J_a^{z}}\right)}\\&+e^{-\frac{1}{4}\beta\left(-4J_a^{xy}+2J_a^{z}+4J_{ab}^{z}+h-2\sqrt{20(J_a^{xy})^2+(J_a^{z})^2+4J_a^{xy}J_a^{z}}\right)}\\&+e^{-\frac{1}{4}\beta\left(-4 J_a^{xy}+2 J_a^{z}+4J_{ab}^{z}+h+2\sqrt{20(J_a^{xy})^2+(J_a^{z})^2+4J_a^{xy}J_a^{z}}\right)},
	\end{aligned}
	\end{equation}
	\begin{equation}
	\begin{aligned}
	Z(\downarrow\downarrow\downarrow,h)=Z(\uparrow\uparrow\uparrow,-h),
	\end{aligned}
	\end{equation}	
	\begin{equation}
	\begin{aligned}
	&Z(\downarrow\uparrow\uparrow,h)=e^{-\frac{1}{4}\beta\left(-12J_a^z-4 J_{ab}^z-13 h\right)}+e^{-\frac{1}{4}\beta\left(4J_a^{xy}-4 J_a^z-4 J_{ab}^z-9 h\right)}\\&+e^{-\frac{1}{4} \beta  \left(-12 J_a^z+4 J_{ab}^z+11 h\right)}+e^{-\frac{1}{4}\beta\left(4J_a^{xy}-4 J_a^z+4 J_{ab}^z+7 h\right)}\\&+e^{-\frac{1}{4}\beta\left(-2J_a^{xy}-4 J_a^z+2 J_{ab}^z+7 h-2 \sqrt{-2 J_{ab}^z J_a^{xy}+9 (J_a^{xy})^2+(J_{ab}^{z})^2}\right)}\\&+e^{-\frac{1}{4} \beta\left(-2J_a^{xy}-4 J_a^z+2 J_{ab}^z+7h+2 \sqrt{-2J_{ab}^zJ_a^{xy}+9(J_a^{xy})^2+(J_{ab}^{z})^2}\right)}\\&+e^{-\frac{1}{4}\beta\left(-2 J_a^{xy}-4 J_a^z-2 J_{ab}^z-9h-2\sqrt{2 J_{ab}^zJ_a^{xy}+9 (J_a^{xy})^2+(J_{ab}^{z})^2}\right)}\\&+e^{-\frac{1}{4}\beta\left(-2 J_a^{\text{xy}}-4J_a^z-2J_{ab}^z-9h+2 \sqrt{2 J_{ab}^z J_a^{xy}+9 (J_a^{xy})^2+(J_{ab}^{z})^2}\right)}\\&+e^{-\frac{1}{4}\beta\left(2J_a^{xy}+2J_a^z-2J_{ab}^z-5h-2\sqrt{2 J_{ab}^zJ_a^{xy}+5(J_a^{xy})^2-2J_{ab}^zJ_a^z+(J_a^{z})^2-2J_a^{xy}J_a^{z}+(J_{ab}^{z})^2}\right)}\\&+e^{-\frac{1}{4}\beta\left(2 J_a^{xy}+2J_a^z-2 J_{ab}^z-5 h+2 \sqrt{2 J_{ab}^z J_a^{xy}+5 (J_a^{xy})^2-2 J_{ab}^z J_a^z+(J_a^{z})^2-2J_a^{xy}J_a^{z}+(J_{ab}^{z})^2}\right)}\\&+e^{-\frac{1}{4}\beta\left(2 J_a^{xy}+2 J_a^z+2 J_{ab}^z+3h-2\sqrt{-2J_{ab}^zJ_a^{xy}+5 (J_a^{xy})^2+2 J_{ab}^z J_a^z+(J_a^{z})^2-2J_a^{xy}J_a^{z}+(J_{ab}^{z})^2}\right)}\\&+e^{-\frac{1}{4}\beta\left(2 J_a^{xy}+2J_a^z+2J_{ab}^z+3h+2\sqrt{-2J_{ab}^zJ_a^{xy}+5(J_a^{xy})^2+2J_{ab}^z J_a^z+(J_a^{z})^2-2 J_a^{xy}J_a^{z}+(J_{ab}^{z})^2}\right)}\\&+e^{-\frac{\beta x_1^1}{4}}+e^{-\frac{\beta x_2^1}{4}}+e^{-\frac{\beta x_3^1}{4}}+e^{-\frac{\beta x_1^2}{4}}+e^{-\frac{\beta x_2^2}{4}}\\&+e^{-\frac{\beta x_3^2}{4}}+e^{-\frac{\beta x_4^2}{4}}+e^{-\frac{\beta x_1^3}{4}}+e^{-\frac{\beta x_2^3}{4}}+e^{-\frac{\beta x_3^3}{4}}\\&+e^{-\frac{\beta x_4^3}{4}}+e^{-\frac{\beta x_1^4}{4}}+e^{-\frac{\beta x_2^4}{4}}+e^{-\frac{\beta x_3^4}{4}}+e^{-\frac{\beta x_4^4}{4}},
	\end{aligned}
	\label{tr}
	\end{equation}	
	\begin{equation}
	\begin{aligned}
	Z(\uparrow\downarrow\downarrow,h)=Z(\downarrow\uparrow\uparrow,-h).
	\end{aligned}
	\end{equation}
	
	In Eq.\ref{tr}, $\{x^1_1, x^1_2, x^1_3\}$ correspond to the roots of a cubic function which is
	
	\begin{equation}
	\left\{
	\begin{array}{rl}
	x^3+&a_1x^2+b_1x+c_1=0\\
	a_1=&-8 J_a^{xy}-12 J_a^z+3 h\\
	b_1=&-16hJ_a^{xy}-24hJ_a^z+64J_a^{xy}J_a^{z}-16(J_a^{xy})^2\\&+48(J_a^{z})^2-16(J_{ab}^{z})^2+3h^2\\
	c_1=&-8 h^2 J_a^{xy}-12 h^2 J_a^z+64 h J_a^{xy}J_a^{z}-16 h (J_a^{xy})^2\\&+48h(J_a^{z})^2+64(J_a^{xy})^2J_a^{z}-128 J_a^{xy}(J_a^{z})^2+128(J_a^{xy})^3\\&-64J_a^{3 z}+h^3+64J_a^z(J_{ab}^{z})^2+16h(J_{ab}^{z})^2
	\end{array}
	\right..
	\label{Cubic function}
	\end{equation}
	
	The analytical roots fo Eq.\ref{Cubic function} can be reached with the general solution to the cubic equation with real coefficients.
	
	Meanwhile, $\{x^2_1, x^2_2, x^2_3, x^2_4\}$, $\{x^3_1, x^3_2, x^3_3, x^3_4\}$ and $\{x^4_1, x^4_2, x^4_3, x^4_4\}$ are the roots of the quartic equations Eqs. \ref{quartic equations01}, \ref{quartic equations02} and \ref{quartic equations03} respectively, which are
	
	\begin{equation}
	\left\{
	\begin{array}{rl}
	x^4+&a_2x^3+b_2x_2+c_2x+d_2=0\\
	a_2=&8J_a^{xy}-12J_a^z+4h\\
	b_2=&24hJ_a^{xy}-36hJ_a^z-64J_a^{xy}J_a^{z}-112(J_a^{xy})^2\\&+48(J_a^{z})^2-16(J_{ab}^{z})^2+6h^2\\
	c_2=&64J_a^z (J_{ab}^{z})^2+24h^2J_a^{xy}-36h^2J_a^z\\&-128hJ_a^{xy}J_a^{z}-224h(J_a^{xy})^2+96h(J_a^{z})^2\\&+832(J_a^{xy})^2J_a^{z}+128J_a^{xy}(J_a^{z})^2-128 (J_a^{xy})^3\\&-64(J_a^{z})^3-32h(J_{ab}^{z})^2+4h^3\\
	d_2=&64hJ_a^z(J_{ab}^{z})^2+512(J_a^{xy})^2 (J_{ab}^{z})^2+8h^3J_a^{xy}\\&-12h^3J_a^z-64h^2J_a^{xy}J_a^{z}-112h^2(J_a^{xy})^2\\&+48h^2(J_a^{z})^2+832h(J_a^{xy})^2J_a^{z}+128hJ_a^{xy}(J_a^{z})^2\\&-128h(J_a^{xy})^3-64h(J_a^{z})^3-1536(J_a^{xy}J_a^{z})^2\\&+1536(J_a^{xy})^4-16h^2(J_{ab}^{z})^2+h^4
	\end{array}
	\right.,
	\label{quartic equations01}
	\end{equation}
	
	\begin{equation}
	\left\{
	\begin{array}{rl}
	x^4+&a_3x_3+b_3x^2+c_3x+d_3=0\\
	a_3=&4J_a^{xy}-8J_a^z+4 J_{ab}^z+20h\\
	b_3=&16J_a^{xy}J_{ab}^z-32J_a^zJ_{ab}^z+60hJ_a^{xy}-120hJ_a^z-32J_a^{xy}J_a^{z}\\&-112(J_a^{xy})^2+16(J_a^{z})^2+60hJ_{ab}^z-16(J_{ab}^{z})^2\\&+150h^2\\
	c_3=&160hJ_a^{xy}J_{ab}^z-320hJ_a^zJ_{ab}^z-64J_a^{xy}(J_{ab}^{z})^2\\&-128(J_a^{xy})^2J_{ab}^z-128 J_{ab}^z
	J_a^{xy}J_a^{z}+64(J_a^{z})^2J_{ab}^z\\&+300h^2J_a^{xy}-600h^2J_a^z-320hJ_a^{xy}J_a^{z}\\&-1120h(J_a^{xy})^2+160h(J_a^{z})^2+576(J_a^{xy})^2J_a^{z}\\&+64J_a^{xy}(J_a^{z})^2+128(J_a^{xy})^3+300h^2J_{ab}^z\\&-160h(J_{ab}^{z})^2-64(J_{ab}^{z})^3+500h^3\\
	d_3=&400h^2J_a^{xy}J_{ab}^z-800h^2J_a^zJ_{ab}^z-320hJ_a^{xy}(J_{ab}^{z})^2\\&-640h(J_a^{xy})^2J_{ab}^z-640 hJ_{ab}^zJ_a^{xy}J_a^{z}+320h(J_a^{z})^2J_{ab}^z\\&-256J_a^{xy}(J_{ab}^{z})^3+256(J_a^{xy})^2(J_{ab}^{z})^2-512(J_a^{xy})^3J_{ab}^z\\&+768J_{ab}^z(J_a^{xy})^2J_a^{z}+256J_{ab}^zJ_a^{xy}(J_a^{z})^2+500h^3J_a^{xy}\\&-1000h^3J_a^z-800h^2J_a^{xy}J_a^{z}-2800h^2(J_a^{xy})^2\\&+400h^2(J_a^{z})^2+2880h(J_a^{xy})^2J_a^{z}+320h J_a^{xy}(J_a^{z})^2\\&+640h(J_a^{xy})^3-512(J_a^{xy})^3J_a^{z}-512(J_a^{xy}J_a^{z})^2\\&+1024(J_a^{xy})^4+500h^3J_{ab}^z-400h^2(J_{ab}^{z})^2\\&-320h(J_{ab}^{z})^3+625h^4
	\end{array}
	\right.,
	\label{quartic equations02}
	\end{equation}
	
	\begin{equation}
	\left\{
	\begin{array}{rl}
	x^4+&a_4x^3+b_4x^2+c_4x+d_4=0\\
	a_4=&4J_a^{xy}-8J_a^z-4J_{ab}^z-12 h\\
	b_4=&-16 J_a^{xy} J_{ab}^z+32 J_a^z J_{ab}^z-36 h J_a^{xy}\\&+72hJ_a^z-32J_a^{xy}J_a^{z}-112 (J_a^{xy})^2\\&+16 (J_a^{z})^2+36 h J_{ab}^z-16(J_{ab}^{z})^2\\&+54 h^2\\
	c_4=&96hJ_a^{xy}J_{ab}^z-192 h J_a^z J_{ab}^z-64 J_a^{xy}(J_{ab}^{z})^2\\&+128 (J_a^{xy})^2 J_{ab}^z+128 J_{ab}^zJ_a^{xy}J_a^{z}-64 (J_a^{z})^2 J_{ab}^z\\&+108 h^2 J_a^{xy}-216 h^2J_a^z+192 hJ_a^{xy}J_a^{z}\\&+672 h (J_a^{xy})^2-96 h (J_a^{z})^2+576 J_a^{xy}J_a^{z}\\&+64 J_a^{xy}(J_a^{z})^2+128 (J_a^{xy})^3-108 h^2J_{ab}^z\\&+96 h(J_{ab}^{z})^2+64 (J_{ab}^{z})^3-108 h^3\\
	d_4=&-144 h^2 J_a^{xy} J_{ab}^z+288 h^2 J_a^z J_{ab}^z+192 hJ_a^{xy} (J_{ab}^{z})^2\\&-384 h (J_a^{xy})^2J_{ab}^z-384 hJ_{ab}^z J_a^{xy}J_a^{z}+192 h (J_a^{z})^2 J_{ab}^z\\&+256 J_a^{xy}(J_{ab}^{z})^3+256 (J_a^{xy})^2 (J_{ab}^{z})^2+512 (J_a^{xy})^3J_{ab}^z\\&-768 J_{ab}^z(J_a^{xy})^2J_a^{z}-256 J_{ab}^zJ_a^{xy}(J_a^{z})^2-108 h^3 J_a^{xy}\\&+216 h^3 J_a^z-288 h^2
	J_a^{xy}J_a^{z}-1008 h^2 (J_a^{xy})^2\\&+144 h^2 (J_a^{z})^2-1728 h (J_a^{xy})^2J_a^{z}-192 h J_a^{xy}(J_a^{z})^2\\&-384 h (J_a^{xy})^3-512 (J_a^{xy})^3J_a^{z}-512 (J_a^{xy}J_a^{z})^2\\&+1024 (J_a^{xy})^4+108 h^3J_{ab}^z-144 h^2 (J_{ab}^{z})^2\\&-192 h (J_{ab}^{z})^3+81 h^4
	\end{array}
	\right..
	\label{quartic equations03}
	\end{equation}
	
	All of them can be solved analytically by applying the general solution to the quartic equation with real coefficients. In practice, it is more convenient to find these roots numerically.
	
	\section{}\label{Eigenvector}
	The hexamer eigenvectors in ordered phases are listed as follows. With Eq.\ref{eigenvector}, we express these eigenvectors as  $\ket{\uparrow\uparrow\uparrow}_{b}\otimes\ket{\downarrow\downarrow\downarrow}_{a}$, for instance, where $\uparrow$ donates $S_{b}^{z}=\frac{1}{2}$ and $\downarrow$ stands for $S_{a}^{z}=-1$. Here are the eigenvectors in the ferrimagnetic phases in Fig.\ref{Fig4}, with the same phase marks in Table.\ref{Tab1}.
	
	For phase IV,
	\begin{equation}
	\begin{aligned}
	\ket{\mathrm{IV}}=&\ket{\uparrow\uparrow\uparrow}_{b} \otimes \ket{\downarrow\downarrow\downarrow}_{a},\\
	E_{\mathrm{IV}}=&-1-3J_{a}^{z}.
	\end{aligned}
	\end{equation}
	
	For phase III,
	\begin{equation}
	\begin{aligned}
	\ket{\mathrm{III}}=&\ket{\uparrow\uparrow\uparrow}_{b}\otimes \frac{1}{\sqrt{3}}(\ket{0\downarrow\downarrow}_{a}+\ket{\downarrow 0\downarrow}_{a}+\ket{\downarrow\downarrow 0}_{a}),\\
	E_{\mathrm{III}}=&-2-J_{a}^{z}-2J_{a}^{xy}.
	\end{aligned}
	\end{equation}

	For phase VII,
	\begin{equation}
	\begin{aligned}
	\ket{\mathrm{VII}}_1=&\ket{\uparrow\uparrow\uparrow}_{b}\otimes \frac{1}{\sqrt{2}}(\ket{\downarrow\downarrow 0}_{a}-\ket{0\downarrow\downarrow}_{a}),\\
	\ket{\mathrm{VII}}_2=&\ket{\uparrow\uparrow\uparrow}_{b}\otimes \frac{1}{\sqrt{2}}(\ket{\downarrow 0\downarrow}_{a}-\ket{0\downarrow\downarrow}_{a}),\\
	E_{\mathrm{VII}}=&-2-J_{a}^{z}+J_{a}^{xy}.
	\end{aligned}
	\end{equation}

	For phase II,
	\begin{equation}
	\begin{aligned}
	\ket{\mathrm{II}}=&\ket{\uparrow\uparrow\uparrow}_{b}\\&\otimes \frac{1}{\sqrt{3}}[\cos\theta_{\mathrm{II}}(\ket{\uparrow\downarrow\downarrow}_{a}+\ket{\downarrow\downarrow\uparrow}_{a}+\ket{\downarrow\uparrow\downarrow}_{a})\\&+\sin\theta_{\mathrm{II}}\left(\ket{00\downarrow}_{a}+\ket{0\downarrow 0}_{a}+\ket{\downarrow 00}_{a}\right)],\\
	E_{\mathrm{II}}=&\frac{1}{2} \left(-2+J_{a}^{z}-2 J_{a}^{xy}-\sqrt{(J_{a}^{z})^2+4 J_{a}^{z} J_{a}^{xy}+20 (J_{a}^{xy})^2}\right),
	\end{aligned}
	\end{equation}
	where $\arctan\theta_{\mathrm{II}}=\frac{J_{a}^{z}+2 J_{a}^{xy}+\sqrt{(J_{a}^{z})^2+4 J_{a}^{z} J_{a}^{xy}+20 (J_{a}^{xy})^2}}{4 J_{a}^{xy}}$.

	For phase VI,
	\begin{equation}
	\begin{aligned}
	\ket{\mathrm{VI}}_1=&\ket{\uparrow\uparrow\uparrow}_{b}\\&\otimes \frac{1}{\sqrt{2}}[ \sin\theta_{\mathrm{VI1}} \left(\ket{0\downarrow 0}_{a}-\ket{00\downarrow}_{a}\right)\\&-\cos\theta_{\mathrm{VI1}}\left(\ket{\downarrow\uparrow\downarrow}_{a}-\ket{\downarrow\downarrow\uparrow}_{a}\right) ],\\
	\ket{\mathrm{VI}}_2=&\ket{\uparrow\uparrow\uparrow}_{b}\\&\otimes \frac{1}{\sqrt{2}}[\cos\theta_{\mathrm{VI2}}\left(\ket{\downarrow 00}_{a}-\ket{0\downarrow 0}_{a}\right)\\&-\sin\theta_{\mathrm{VI2}}\left(\ket{\uparrow\downarrow\downarrow}_{a}-\ket{\downarrow\uparrow\downarrow}_{a}\right)],\\
	E_{\mathrm{VI}}=&\frac{1}{2} \left(-2+J_{a}^{z}+J_{a}^{xy}-\sqrt{(J_{a}^{z})^2-2 J_{a}^{z} J_{a}^{xy}+5 (J_{a}^{xy})^2}\right),
	\end{aligned}
	\end{equation}
	where $\arctan\theta_{\mathrm{VI1}}=\frac{J_{a}^{z}-J_{a}^{xy}+\sqrt{(J_{a}^{z})^2-2 J_{a}^{z} J_{a}^{xy}+5 (J_{a}^{xy})^2}}{2 J_{a}^{xy}}$ and $\arctan\theta_{\mathrm{VI2}}=\frac{-J_{a}^{z}+J_{a}^{xy}+\sqrt{(J_{a}^{z})^2-2 J_{a}^{z} J_{a}^{xy}+5 (J_{a}^{xy})^2}}{2 J_{a}^{xy}}$.
	
	\begin{figure}
		
		\centering
		{\subfigure[]{
				\includegraphics[height=3cm]{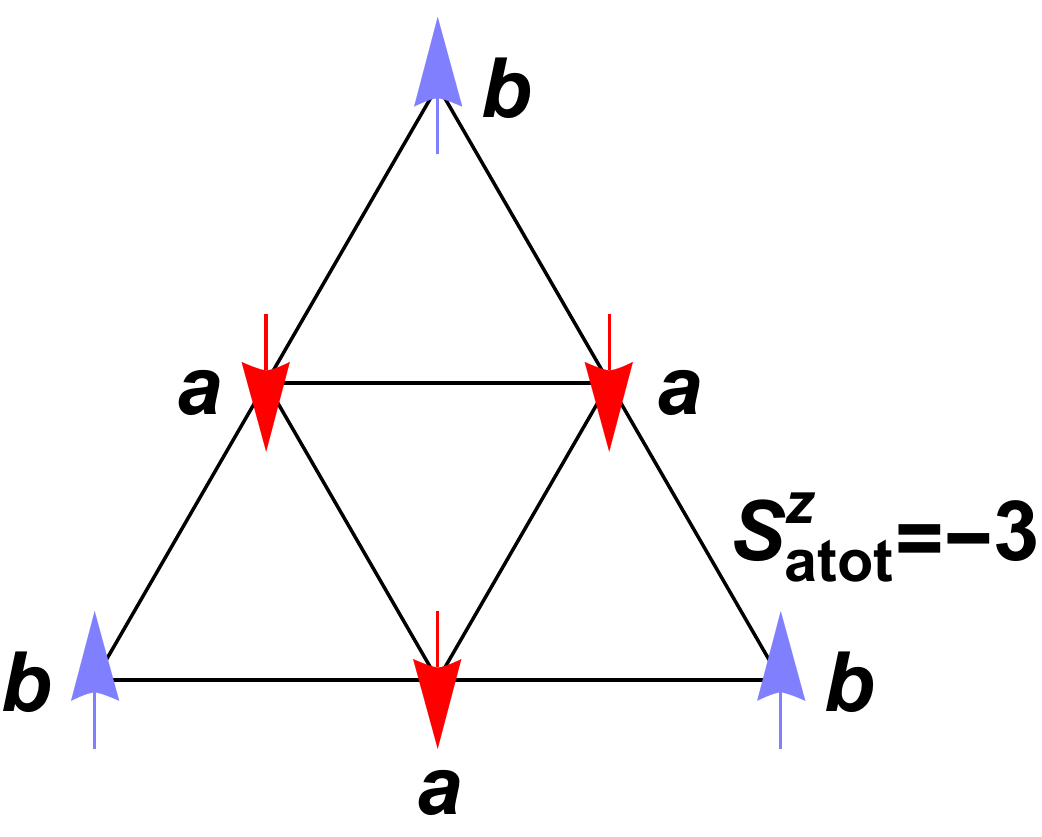}}
			\label{Fig12a}}
		{\subfigure[]{
				\includegraphics[height=3cm]{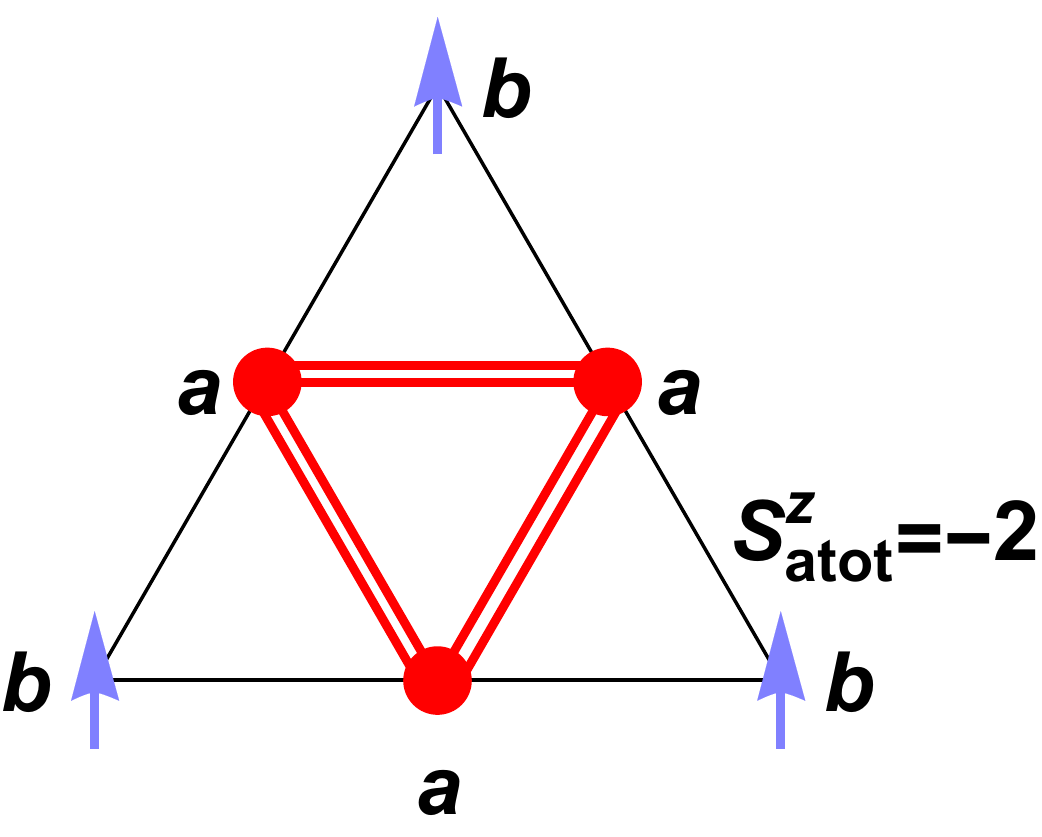}}
			\label{Fig12b}}
		{\subfigure[]{
				\includegraphics[height=3cm]{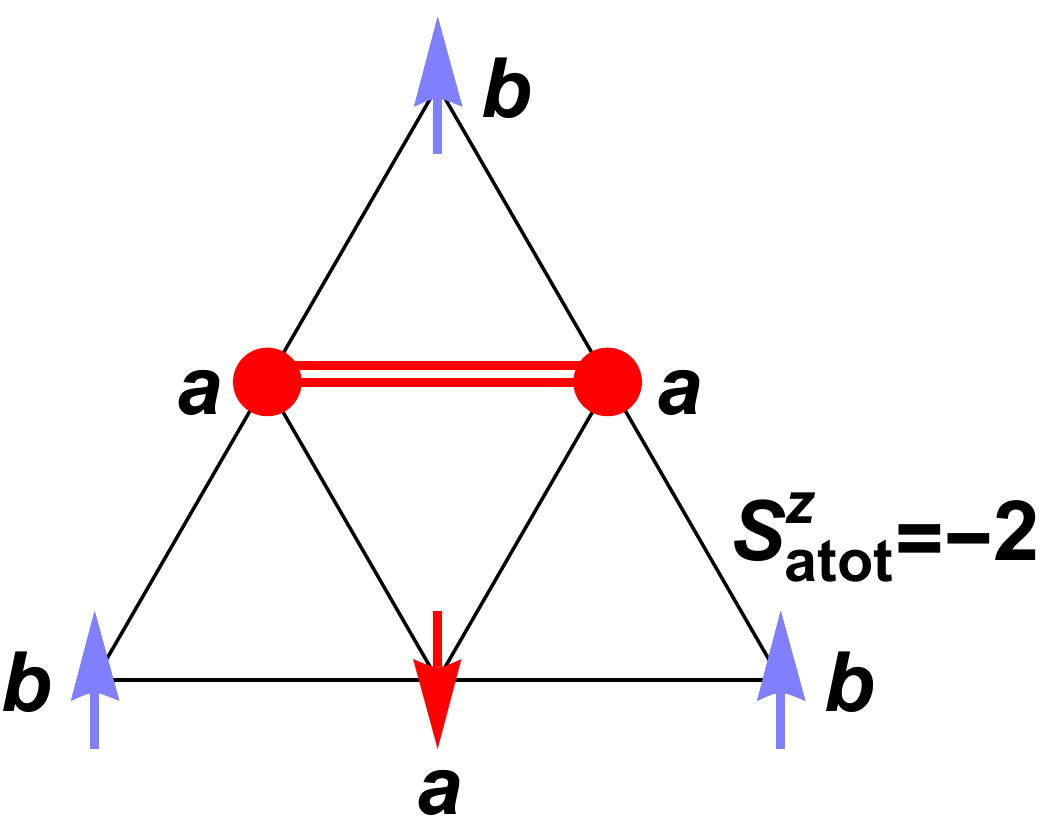}}
			\label{Fig12c}}
		{\subfigure[]{
				\includegraphics[height=3cm]{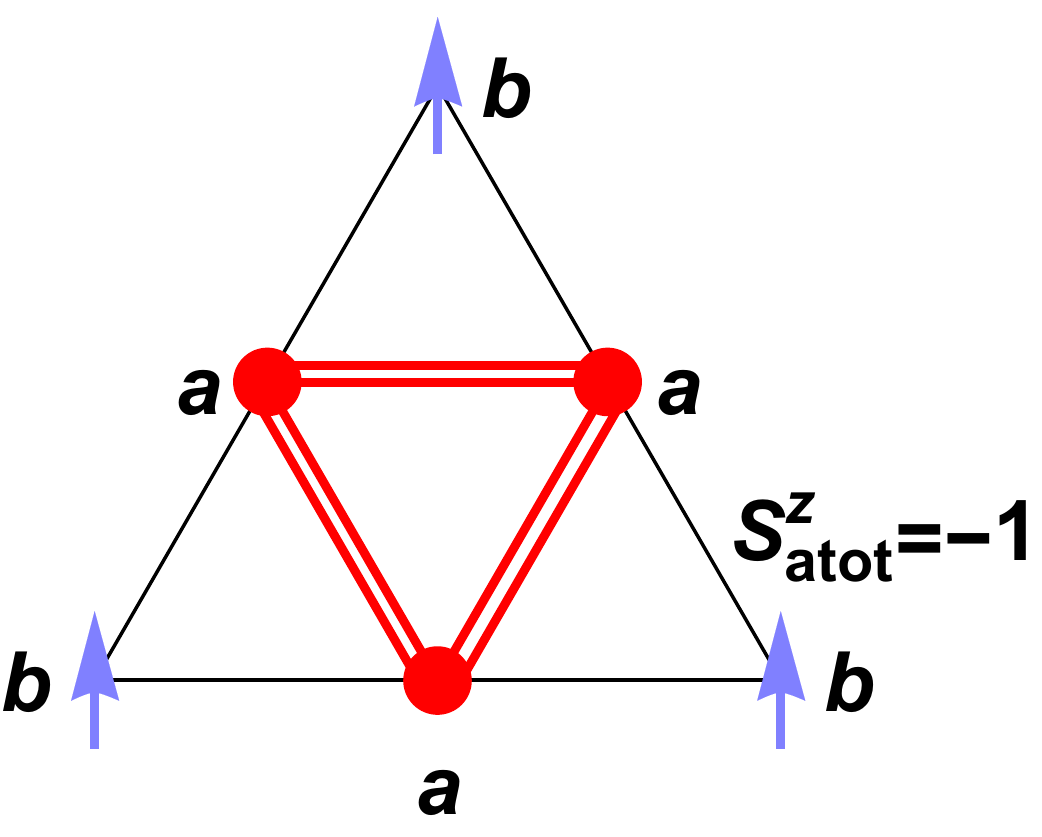}}
			\label{Fig12d}}
		{\subfigure[]{
				\includegraphics[height=3cm]{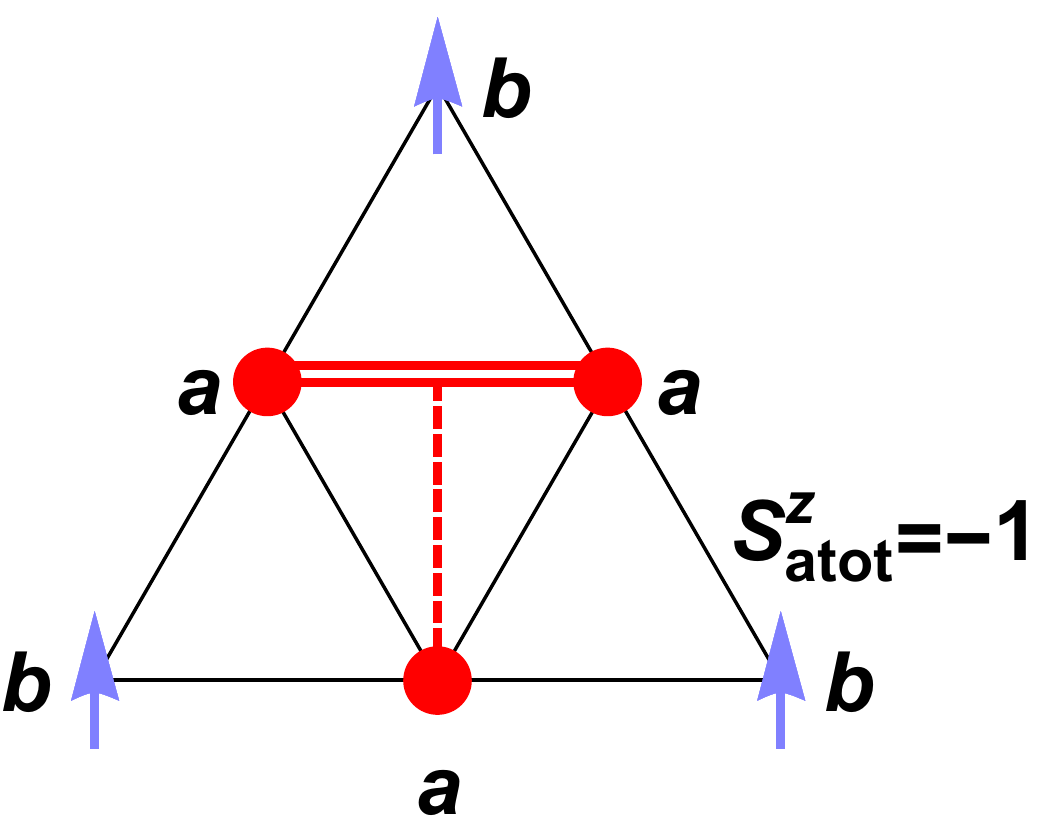}}
			\label{Fig12e}}
		\caption{We plot the schematic diagrams of the spin configuration of each hexamer when there is no external field (Fig.\ref{Fig4}). For instance, the arrow $\uparrow$ represents $S=1$ or $S=1/2$. And the red (thick or dashed) lines are the entangled relationships of the a-spins. Fig.\ref{Fig12}(a) is the spin configuration when the model is in phase IV in Fig.\ref{Fig4}. And Fig.\ref{Fig12}(b) and (d) denote the trimerized states in phase II and phase III respectively. Fig.\ref{Fig12}(c) shows the ordinary dimerized state in phase VII while Fig.\ref{Fig12}(e) presents the anisotropic trimerized state in phase VI. The anisotropic trimerized state can be viewed as a two-step dimerizing. And the thick line is for the first step dimerizing and the dashed line denotes the second step dimerizing.}
		\label{Fig12}
	\end{figure}
	
	When $h\neq0$, we give the eigenvectors, ground state energy of each hexamer and magnetization of each unit cell in those ordered phases in Fig.\ref{Fig6}b (phases I, II, III, and IV in Table.\ref{Tab3}).
	
	\begin{equation}
	\begin{aligned}
	\ket{\mathrm{I}}=&\ket{\uparrow\uparrow\uparrow}_{b}\otimes \frac{1}{\sqrt{6}}(\ket{0\uparrow\downarrow }_{a}+\ket{\uparrow\downarrow 0}_{a}+\ket{\downarrow 0\uparrow}_{a}\\&-\ket{\uparrow 0\downarrow}_{a}-\ket{0\downarrow\uparrow}_{a}-\ket{\downarrow\uparrow 0}_{a} ),\\
	E_{\mathrm{I}}=&\frac{3}{4} \left(h-4J_{a}^{z}\right),\\
	m_{\mathrm{I}}=&\frac{3}{2}.
	\end{aligned}
	\end{equation}
	
	Each a-trimer in phase I is in a singlet trimerized state. 
	
	\begin{equation}
	\begin{aligned}
	\ket{\mathrm{II}}=&\ket{\uparrow\uparrow\uparrow}_{b}\otimes \frac{1}{\sqrt{3}}(\ket{\uparrow\uparrow\downarrow }_{a}-\ket{0\uparrow 0}_{a}+\ket{\downarrow \uparrow\uparrow}_{a} ),\\
	E_{\mathrm{II}}=&\frac{1}{4} \left(4-7h+8J_{a}^{z}\right),\\
	m_{\mathrm{II}}=&\frac{7}{2}.
	\end{aligned}
	\end{equation}

	\begin{equation}
	\begin{aligned}
	\ket{\mathrm{III}}=&\ket{\uparrow\uparrow\uparrow}_{b}\otimes \frac{1}{\sqrt{2}}(\ket{0\uparrow\uparrow }_{a}-\ket{\uparrow\uparrow 0}_{a} ),\\
	m_{\mathrm{III}}=&\frac{11}{2}.
	\end{aligned}
	\end{equation}
	
	In phase II and phase III, the a-trimers are in the dimerized state. We don't give their $C_3$ symmetry counterparts here, which cause the macroscopic degeneracy.
	
	\begin{equation}
	\begin{aligned}
	\ket{\mathrm{IV}}=&\ket{\uparrow\uparrow\uparrow}_{b}\otimes\ket{\uparrow\uparrow\uparrow }_{a},\\
	E_{\mathrm{IV}}=&\frac{3}{4} \left(-4+5h+4J_{a}^{z}\right),\\
	m_{\mathrm{IV}}=&\frac{15}{2}.
	\end{aligned}
	\end{equation}
	
	The a-trimers in phase IV are in the classical state. The spin-$1$ TKL model meets its saturation magnetization in phase IV 
	
	As $h$ increases, the spin-$1$ TKL model changes from phase I to phase IV(see Fig.\ref{Fig6}(b)). These phase transitions correspond to that each a-trimer develops from the trimerized state, to dimerized state, and finally to the classical state. It is also responsible for the stable magnetization plateaus (see Fig.\ref{Fig11}). 
	
		\begin{figure}
		\centering
		{\subfigure[]{
				\includegraphics[height=3cm]{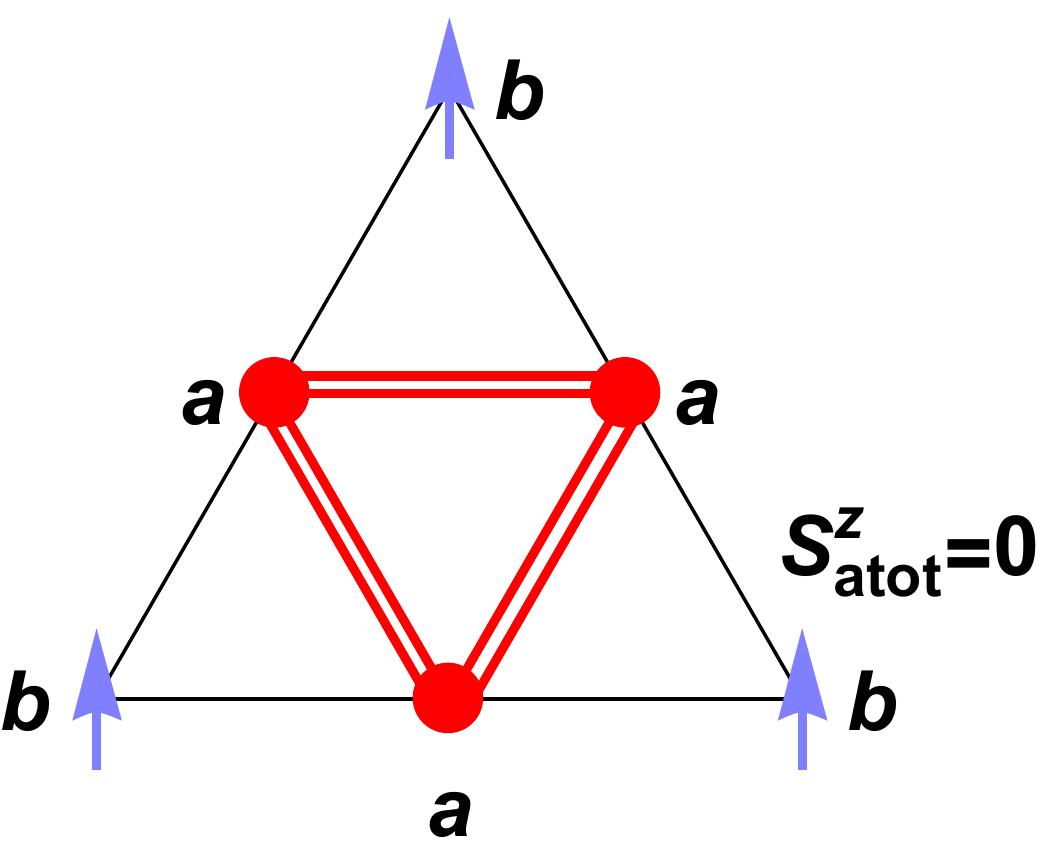}}
			\label{Fig13a}}
		{\subfigure[]{
				\includegraphics[height=3cm]{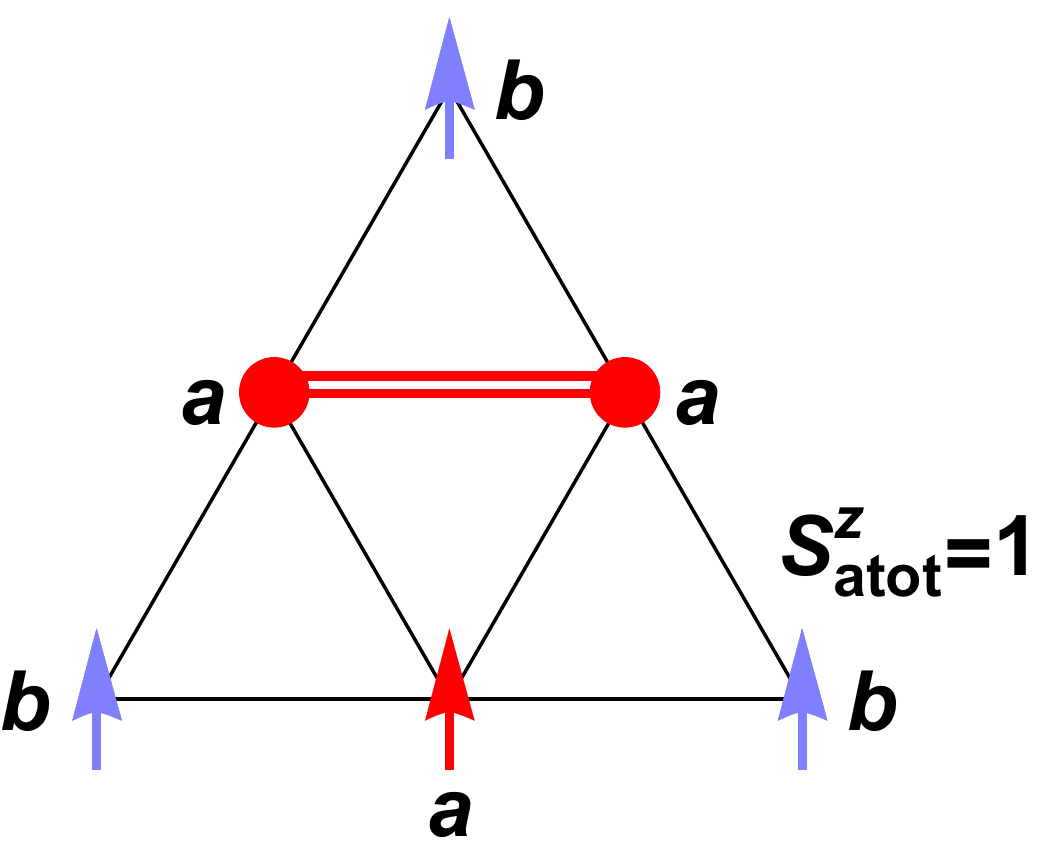}}
			\label{Fig13b}}
		{\subfigure[]{
				\includegraphics[height=3cm]{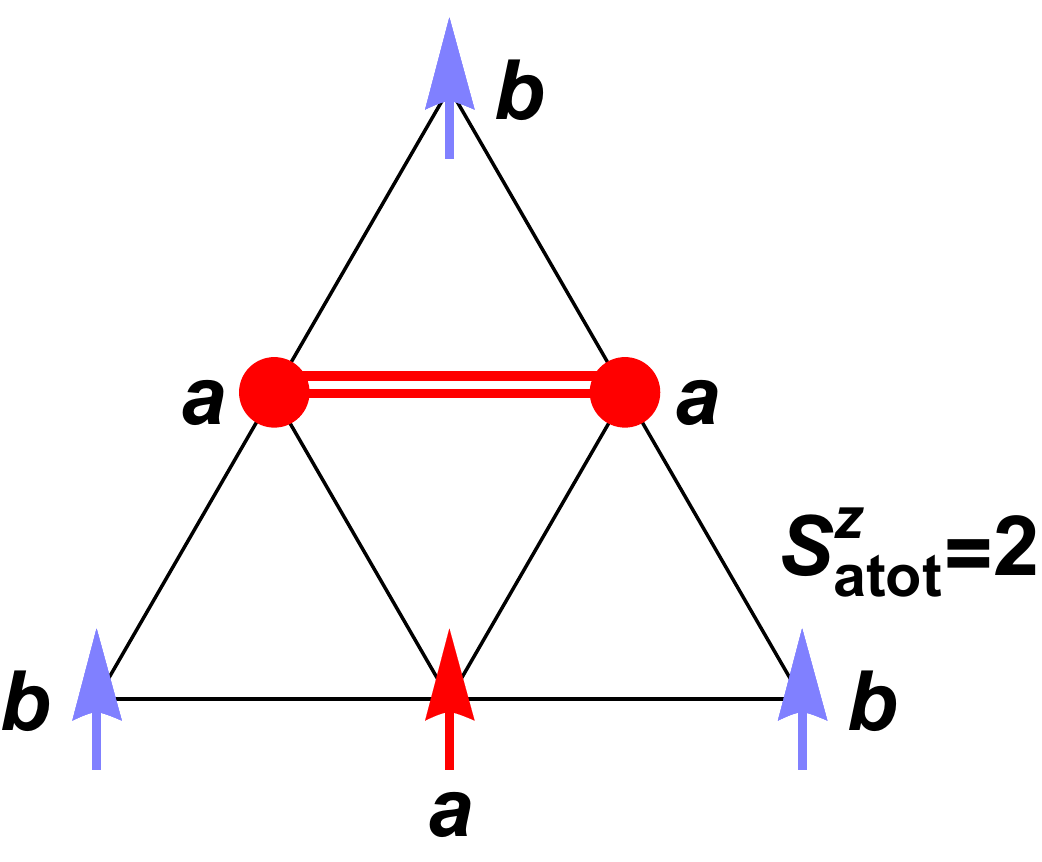}}
			\label{Fig13c}}
		{\subfigure[]{
				\includegraphics[height=3cm]{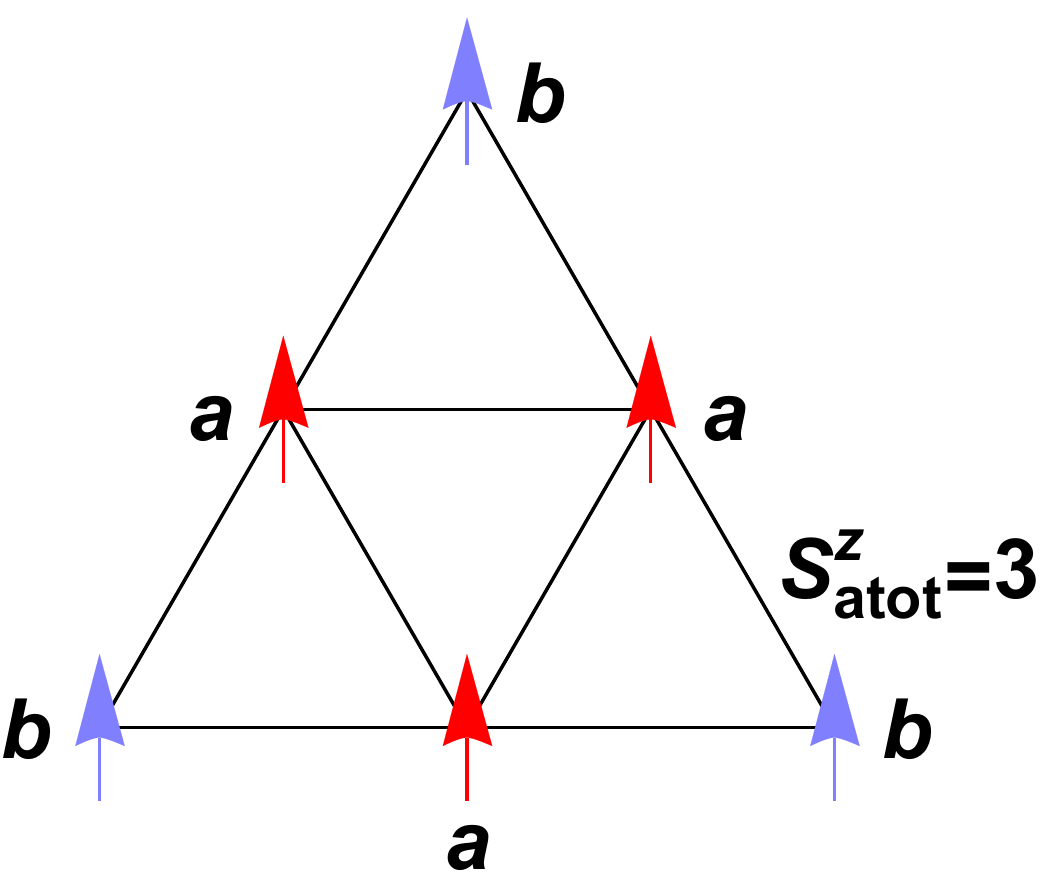}}
			\label{Fig13d}}
		\caption{We plot the schematic diagrams of the spin configuration of each hexamer in the ordered phases when $h\neq0$ (Fig.\ref{Fig6}). Also, the arrow $\uparrow$ represents $S=1$ or $S=1/2$ and the red (thick or dashed) lines are the entangled relationships of the a-spins. Fig.\ref{Fig13}(a) is the spin configuration when the model is in phase I in Fig.\ref{Fig6}, which is a singlet trimerized state. And Fig.\ref{Fig13}(b) and (c) denote the dimerized states in phase II and phase III respectively. Finally, Fig.\ref{Fig13}(d) presents the classical state in phase IV. As $h$ increases, the spin arrangement of each hexamer changes from Fig.\ref{Fig13}(a), to Fig.\ref{Fig13}(b), to Fig.\ref{Fig13}(c), and finally to Fig.\ref{Fig13}(d), which is a step-by-step disentangled process.}
		\label{Fig13}
	\end{figure} 

	\begin{acknowledgments}
	We thank E. W. Carlson, Y. L. Loh, Y. -R. Shu, M. Lake and N. Raper for useful discussions.  This project is supported by NKRDPC-2017YFA0206203, NSFC-11574404, NSFC-11275279, NSFG-2015A030313176, Special Program for Applied Research on Super Computation of the NSFC-Guangdong Joint Fund, Three Big Constructions—Supercomputing Application Cultivation Projects,
	 and the Leading Talent Program of Guangdong Special Projects.
	\end{acknowledgments}

	\bibliography{tkls1}
	\bibliographystyle{apsrev4-1}

\end{document}